\documentclass[amsmath,amssymb,aps,prb,twocolumn]{revtex4-2}
\usepackage{graphicx}
\usepackage{dcolumn}
\usepackage{bm}
\usepackage[colorlinks,allcolors=blue]{hyperref} 
\usepackage{hyperref}
\usepackage{amsmath}
\usepackage{amsfonts}
\usepackage{amssymb}
\usepackage{natbib}
\usepackage{textcomp,gensymb}
\usepackage{color}

\newcommand{\ket}[1]{\mbox{$|#1\rangle$}}
\begin{document}
\title{Error analysis in large area multi-Raman pulse atom interferometry due to
undesired spontaneous decay}
\author{Philip Chrostoski$^{\ast}$}
\author{Scott Bisson$^{\ast}$}
\author{David Farley$^{\ast}$}
\author{Frank A. Narducci$^{\dagger}$}
\author{Daniel Soh$^{\ddagger}$}
\affiliation{$^{\ast}$Sandia National Laboratories, Livermore, California 94550, USA}
\affiliation{$^{\dagger}$Department of Physics, Naval Postgraduate School, Monterey, California 93945, USA}
\affiliation{$^{\ddagger}$Wyant College of Optical Sciences, the University of Arizona, Tucson, Arizona 85721, USA}

\begin{abstract}

Despite the fact that atom interferometry has been a successful application of quantum sensing, a major topic of interest is the further improvement of the sensitivity of these devices.
In particular, the area enclosed by the interferometer (which controls the sensitivity) can be increased by providing a larger momentum kick to the atom cloud, increasing the extent of the momentum axis. One such atom optics technique involves increasing the number of central $\pi-$Raman  pulses. This technique, while providing the prerequisite additional momentum boost, also causes the atom to remain  in the intermediate high energy state for longer periods of time. This additional length of time is often neglected in many treatments due to the adiabatic elimination of the higher energy state enabled by the large optical detuning. The increased time in the intermediate high energy state results in a higher probability of undesired spontaneous decay and a loss of quantum information, thereby adding error to the atom interferometer. In this work, we consider an open quantum system using the Lindblad master equation to devise a model for the atomic state dynamics that includes the undesired spontaneous decay from the intermediate high energy
state. We formulate an error figure of merit to analyze limitations of an atom interferometer configured for acceleration measurements. Our theoretical results show the error figure of merit will be dominated by a $N_{R}^{-2}$ scaling factor for low numbers of $\pi-$Raman pulses, but will be dominated by a monotonic increase in error for high number of $\pi-$Raman pulses. We determined the number of $\pi$-Raman pulses that accomplishes maximal momentum transfer with a the minimal error, depending on major system parameters.

\end{abstract}
\date{\today }
\maketitle

\section{Introduction}

Atom interferometry has developed into an active field over the last
decade. Atom interferometers using either cold thermal atoms or Bose-Einstein condensates (BECs) have led to many applications including gravimeters \cite{HSD13,CFB11,ACC99}, gyroscopes
\cite{STK11,GCL09,DSK06}, magnetometers \cite{DN08}, photon recoil
determination \cite{BCK11}, and tests of the foundations of general relativity
\cite{HM13}. Though atom interferometry has
allowed for ultra precise measurements, the fundamental limitation is set by the small momentum separation, which will have a smaller sensitivity. Increasing the overall momentum separation will improve the sensitivity, which is necessary to make ultra precise measurements and practical application. The area enclosed in phase space is directly proportional to the momentum separation. Employing atom optics that will generate large momentum transfer (LMT) will then lead to enhancements in the sensitivity \cite{SCT94,MSK00}. There have been several approaches to LMT atom optics such as
applying multi-photon pulses to drive Bragg transitions \cite{MCL08,CKC11,KCK12}, Bloch
oscillations in an optical lattice \cite{CKN09,MCH09}, stimulated Raman
transitions \cite{MSK00,AS14}, composite Raman transitions \cite{BKK13,BAT15},
stimulated Raman adiabatic rapid passage (STIRAP) \cite{WYC94},
frequency-swept Raman adiabatic rapid passage (ARP) \cite{KBK15}, and clock state transitions \cite{RWN20}.

LMT using multiple-Raman pulse techniques has been shown to generate momentum
transfers as large as $6\hbar k_{\mathrm{eff}}$ \cite{MSK00,BAT15}, $18\hbar k_{\mathrm{eff}}$ (using composite Raman pulses) \cite{BKK13}, and $30\hbar k_{\mathrm{eff}}$ (combining with frequency swept adiabatic passage) \cite{KBK15} and even up to 102$\hbar k_{eff}$ momentum transfer with unresolved interference measurement \cite{KAS11} compared to $2\hbar k_{\mathrm{eff}}$ from a non-LMT standard method. One issue that is believed to reduce the performance of atom interferometers using these LMT techniques is spontaneous emission. To the best of our knowledge, a theoretical analysis of the error due to an increase in spontaneous decay of the intermediate high-energy state has not been done until now. In this work, we consider an open quantum system atom interferometer and apply the Lindblad master equation in the Schr\"{o}dinger picture to formulate the new dynamics of the atomic quantum states (section (\ref{sec_dynamics})). The open quantum system will have an undesired spontaneous decay causing loss of quantum information. In section (\ref{sec_error}), we derive the variance in the measurement of acceleration by the atom interferometer, whose contributions come from mainly two error sources: the measurement uncertainty and the undesired spontaneous decay. We next combined the variance (AC fluctuation) and the mean value deviation (DC offset) into an error figure of merit (FOM) to analyze the limitations of LMT atom interferometry using multiple Raman pulses.

In section (\ref{sec_results}) we present the error analysis of an open quantum system atom interferometer acceleration measurement affected by undesired spontaneous decay. We first simulate the LMT atom interferometer acceleration measurement by numerically solving the open quantum system dynamics using the Runge-Kutta 4th order method \cite{PTV07}. Our simulations are run multiple times for each LMT amount giving statistics on the measured acceleration deviation allowing us to calculate the variance for three different cases (section (\ref{sec_PhotonCountResults})): a lossless (i.e., no spontaneous decay) case, the quantum information loss being a constant (expected for atom amounts on to order of $10^{6}$), and lastly considering the quantum information loss to be a random variable in the variance calculation.

In section (\ref{sec_PhotonCountResults}) we examine the FOM and incorporate photon counting error to determine the pulse amount with minimal error. We then look at the variance term of the FOM (Eqn. (\ref{FOM})) alone to see how it affects where the minimum FOM occurs when compared to the effect of the DC offset term. We next consider that increasing the large single-photon detuning of the high-energy intermediate state is believed to limit the amount of spontaneous decay. Experiments have used single-photon detuning amounts on the order of $2-4$ $\mathrm{GHz}$ \cite{MSK00,BKK13,KBK15}. On the other hand, the effect of the two-photon detuning of the low-energy excited state is also an important factor in the performance of the atom interferometer. With this in mind, we study how the single- and two-photon detunings affect how much LMT can be achieved when spontaneous emission is included (section  \ref{sec_DeltaResults} and \ref{sec_deltaResults}).

\section{Model and calculations \label{sec_model}}

\subsection{Open quantum system: spontaneous decay dynamics \label{sec_dynamics}}

To understand the difference between the  multiple Raman light-pulse sequence difference and the
usual Mach-Zehnder atom interferometer, let us consider a
three-level atomic system and summarize the system dynamics
which has been worked out before \cite{JBF03,AVR15,BVE16,CF17,PB97}. The Hilbert
space will be made up of three ``real'' orthonormal internal quantum states: $\left\vert
g\right\rangle $ (ground), $\left\vert e\right\rangle $ (excited), $\left\vert
i\right\rangle $ (intermediate) and external momentum states: $\left\vert \mathbf{p}%
_{g}\right\rangle ,\left\vert \mathbf{p}_{e}\right\rangle ,\left\vert
\mathbf{p}_{i}\right\rangle $. Due to the momentum exchange between photons and atoms, and the associated energy change as a result, the Hermitian part of the system is described by the unperturbed Hamiltonian
\begin{align}
\widehat{H}_{0}=\hbar \omega _{ge}\left\vert e\right\rangle \left\langle
e\right\vert +\hbar \omega _{gi}\left\vert i\right\rangle \left\langle
i\right\vert.
\end{align}
Here $\hbar \omega _{gi}$ is the energy between $\left\vert g\right\rangle $
and $\left\vert i\right\rangle $ and $\hbar \omega _{ge}$ is the energy
difference between $\left\vert g\right\rangle$ and $\left\vert
e\right\rangle$. The non-Hermitian part is described through the spontaneous decay to an additional state we add by hand which we refer to as the loss state $\vert l\rangle$. This state is made up of all the sub-levels that are not $\left\vert g\right\rangle$ and $\left\vert e\right\rangle$ (described further below). With the system defined, we can write the interaction Hamiltonian in the following way \cite{PB97}
\begin{align}
&\widehat{H}_{\text{int}}  =\nonumber\\
&  \hbar\Omega_{1}^{\ast}e^{i(\mathbf{k}_{1}\cdot\mathbf{x}_{g}-\omega_{1}%
t)}\left\vert i\right\rangle \left\langle g\right\vert +\hbar\Omega_{2}^{\ast
}e^{i(\mathbf{k}_{2}\cdot\mathbf{x}_{e}-\omega_{2}t)}\left\vert i\right\rangle
\left\langle e\right\vert +h.c.
\end{align}
Here $\mathbf{x}_{g,e}$ represent the positions of the particle's $\vert g\rangle$ and $\left\vert e \right\rangle$ states respectively, $\hbar$ is the reduced Planck's constant, $\omega_{1,2}$ are the Raman beam laser frequencies, $\mathbf{k}_{1,2}$ are the wave vectors, and $\Omega_{1,2}$ are the single-photon Rabi frequencies. Also, `h.c.' stands for complex Hermite conjugate of operator terms.
\begin{figure}[ptb]
	\centering
	\includegraphics[width=\columnwidth]{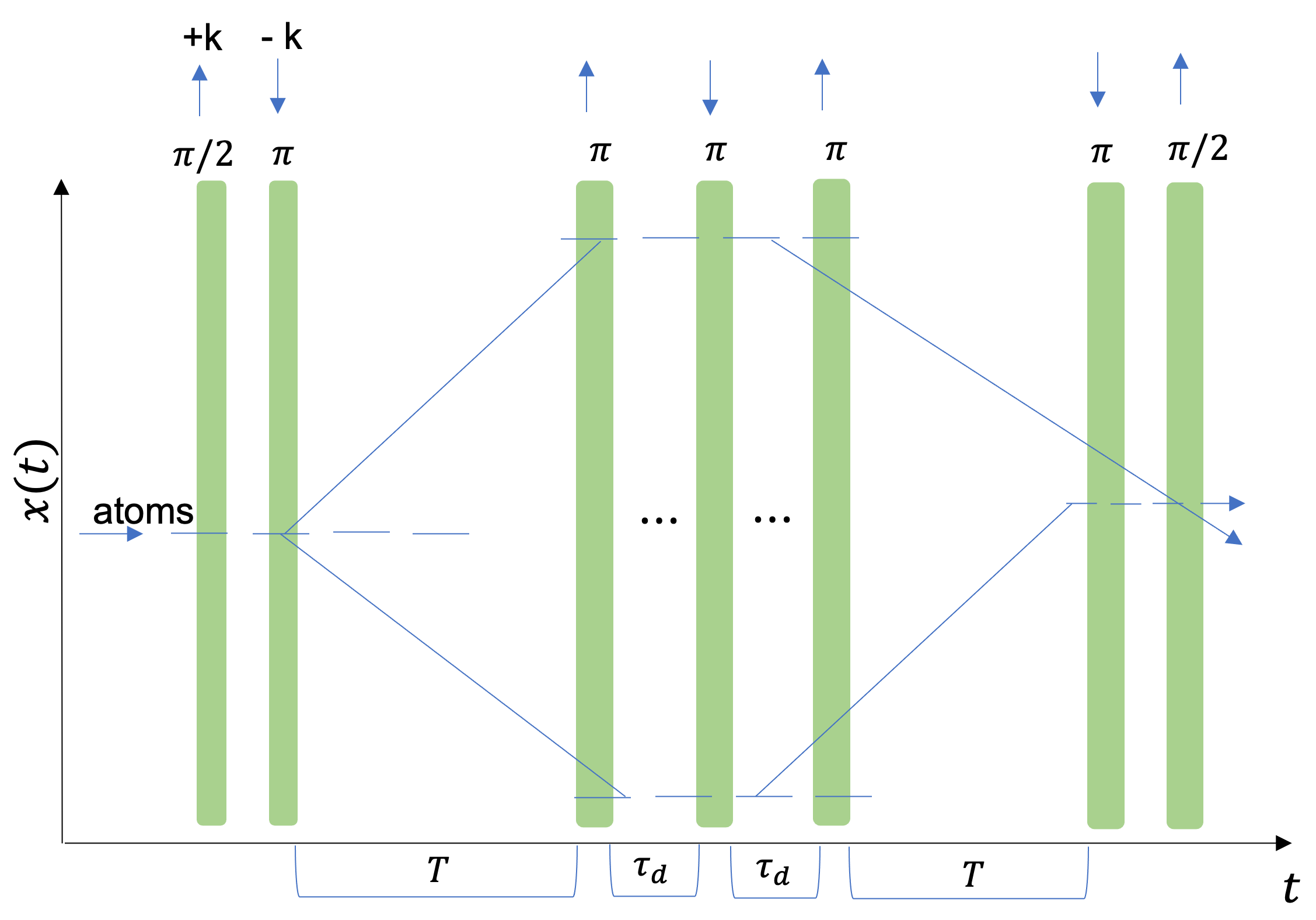}\caption{Recoil diagram for a $N_{R}=3$ ($6\hbar k_\mathrm{eff}$) LMT multiple light pulse sequence. The sequential pulses occur a time $\tau_{d}$ after each other. The number of sequential $\pi$-pulses after and before the $\pi/2$ pulses will increase by $(N_{R}-1)/2$ for higher LMT. Here $T$ is the free evolution time.}%
	\label{fig1_LMTmodel}%
\end{figure}

We now consider how the pulse sequence will change for the LMT multiple-Raman pulse system. The LMT comes from the increase in the number of central $\pi$-pulses of the system which can be seen in Fig. \ref{fig1_LMTmodel}. The $\pi$ pulse in rapid succession after the initial $\pi/2$ pulse increases the initial momentum separation whereas the final $\pi$ pulse before makes sure the atom cloud recombines with the final $\pi/2$ pulse \cite{MSK00}. 

\begin{figure}[ptb]
	\centering
	\includegraphics[width=\columnwidth]{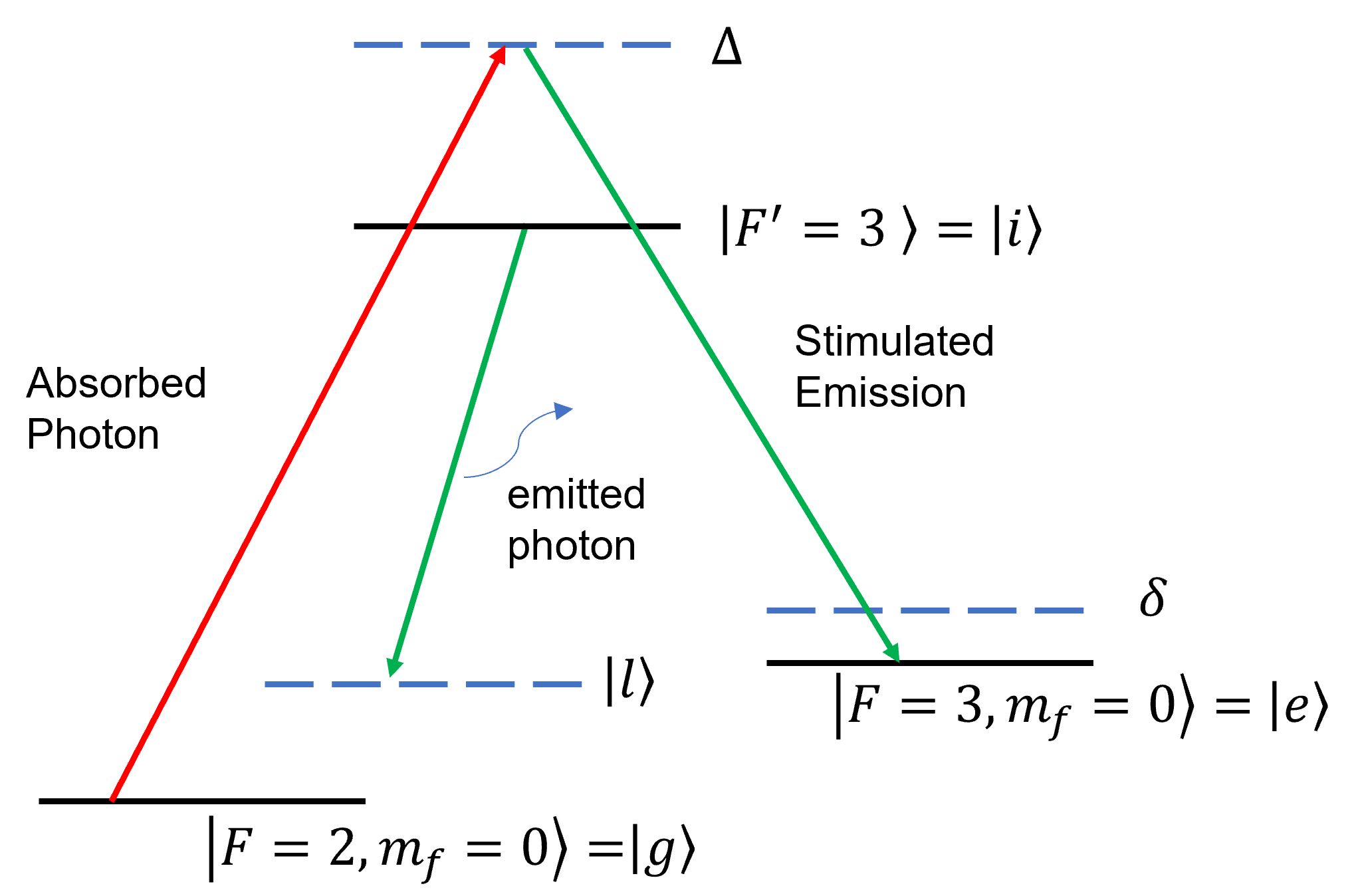}\caption{Updated energy diagram of the three-level system. $\Delta$ and $\delta$ are the single- and two-photon detunings of the high-energy intermediate and excited states respectively. Spontaneous decay will drop the atom into a loss state, $\left\vert l\right\rangle$, made up of all the sub-levels that are not the $\left\vert g\right\rangle$ and $\left\vert e\right\rangle$ states. This decay will emit a photon of an indeterminate mode giving pure loss.}%
	\label{fig2_EnergyDiagram}%
\end{figure}

The increase in the amount of Raman pulses will cause there to be a higher probability to occupy the
$\left\vert i\right\rangle$ state which can undergo spontaneous decay. The spontaneous decay of the $\left\vert i\right\rangle $ state will emit a photon of an indeterminate mode making the information
irretrievable (Fig. \ref{fig2_EnergyDiagram}). Due to this Markovian process, we can apply the Lindblad
master equation to determine the modified dynamics. We will consider the Schr\"{o}dinger picture and the density matrix $\rho$ which will give us \cite{GL76,DM20,CSE22}
\begin{equation}
\overset{\cdot}{\rho}=-\frac{i}{\hbar}\left[H,\rho\right]  +\sum_{k}\left(L_{k}\rho L_{k}^{\dagger}-\frac{1}{2} L_{k}^{\dagger}L_{k}\rho - \frac{1}{2} \rho L_{k}^{\dagger}L_{k}\right) .
\end{equation}
The jump operators from state $\left\vert i\right\rangle $ to $\left\vert g\right\rangle$, $\left\vert e\right\rangle$, and $\left\vert l\right\rangle$ are  given by $\widehat{g} = |g \rangle \langle i|$, $\widehat{e} = |e \rangle \langle i|$, and $\widehat{l} = |l \rangle \langle i|$.
The Lindblad operators will be $L_{g}=\sqrt{\gamma_{g}}g$, $L_{e}=\sqrt{\gamma_{e}}e$, and $L_{l}=\sqrt{\gamma_{l}}l$ where $\gamma_{g,e,l}$ are the spontaneous decay rates from state $\left\vert i\right\rangle$ to $\left\vert g\right\rangle$, $\left\vert e\right\rangle$, and $\left\vert l\right\rangle$ respectively.  We will also note that we are considering a system of Rubidium-85 ($^{85}$Rb). This system will consist of a $\left\vert F = 2, m_{f}=0 \right\rangle$ ground state, a $\left\vert F = 3, m_{f}=0 \right\rangle$ excited state, and a $\left\vert F^{\prime} = 3 \right\rangle$ intermediate state. In this context, ``excited'' refers to the electronic ground state of higher energy. Depending on the polarization of the Raman beam, there will be a probability of occupying either the $m_{f}+1$ or $m_{f}=0$ state of $F^{\prime}=3$. 
The spontaneous decay can then be determined by knowing the dipole moments between the sub-levels and calculating the Einstein A coefficients. With knowledge of the magnitude of the dipole moments from the $^{85}$Rb D-line data \cite{DAS21}, it can be seen that the spontaneous decay will be dominated by a decay process in which an emited a photon of an indeterminate mode ends up in the loss state made up of all the sub-levels that are not the excited or ground state. It is also important to note that the dipole moments between the excited and loss state and the polarizations will strongly suppress (even forbid) transitions between the excited state and the loss state. We will make the approximation that the $L_{k}$$\rho$$L_{k}^{\dagger}$ term can be neglected due to depumping and fast spontaneous decay and reduce our master equation system dynamics to consider only the dominating decay branch giving us \cite{CSE22,TDL19,RS12},
\begin{equation}
\overset{\cdot}{\rho}=-\frac{i}{\hbar}\left[H,\rho\right]  -\frac{1}{2}\left(L_{l}^{\dagger}L_{l}\rho + \rho L_{l}^{\dagger}L_{l}\right).
\label{Eq:Lindblad}
\end{equation}

With the reduced master equation, we can follow the general approach of Chatterjee \textit{et al} to get an effective Hamiltonian \cite{CSE22}, 
\begin{equation}
\widehat{H}_{\text{eff}} = \widehat{H}_{0} + \widehat{H}_{\text{int}} - \frac{i\hbar}{2}L_{l}^{\dagger}L_{l}.
\end{equation}
The Hamiltonian is made up of a Hermitian part given by $\widehat{H}_{0} + \widehat{H}_{\text{int}}$ and a non-Hermitian part from the Lindbladian. With our Hamiltonian we can use the Schrodinger equation to determine the three-level system dynamics using a quantum state given as 
\begin{equation}
    |\psi(t)\rangle = c_g(t) |g\rangle + c_e (t) |e\rangle + c_i (t) |i \rangle,
\end{equation}
and write the solution in terms of the amplitudes (which are related to the density matrix elements) $c_g(t), c_e(t),$ and $c_i(t)$. Then, we obtain the following
\begin{widetext}
\begin{subequations}
\begin{eqnarray}
i\hbar\overset{\cdot}{c}_{g}&=&\hbar\Omega_{1}e^{i(-\mathbf{k}_{1}\cdot
x_{g}+\omega_{1}t)}c_{i}, \\
i\hbar\overset{\cdot}{c}_{e}&=&\hbar\Omega_{2}e^{i(-\mathbf{k}_{2}\cdot
x_{e}+\omega_{2}t)}c_{i}+\hbar\omega_{ge}c_{e},\\
i\hbar\overset{\cdot}{c}_{i}&=&\hbar\Omega_{1}^{\ast}e^{i(\mathbf{k}_{1}\cdot\mathbf{x}%
_{g}-\omega_{1}t)}{c}_{g} + \hbar\Omega_{2}^{\ast}e^{i(\mathbf{k}_{2}\cdot\mathbf{x}_{e}-\omega_{2}t)}{c}_{e}+\hbar\omega_{gi}c_{i}-i\hbar
\frac{\gamma_{l}}{2}c_{i},
\end{eqnarray}
\end{subequations}
\end{widetext}

We next move into the rotating frame and use the substitutions 
\begin{subequations}
\begin{eqnarray}
    c_{g}&=&\widetilde{c}_{g}\\
    c_{e}&=&\widetilde{c}_{e}e^{-i\omega_{ge}t}\\
    c_{i}&=&\widetilde{c}_{i}e^{-i\omega_{gi}t},
\end{eqnarray}
\end{subequations}
(where the tilde denotes the variable in the rotating frame) to help simplify solving this set of coupled differential equations. In standard treatments, the intermediate state is adiabatically eliminated by making the assumption that $\tilde{c}_{g}$ and $\tilde{c}_{e}$  varying slowly compared to the very fast dynamics and decay of $\tilde{c}_{i}$ \cite{BPM07,FVS20}. However, since we are considering the loss of information due to spontaneous emission into a loss state, we need to properly account for spontaneous emission, as either long Raman pulses or multiple $\pi$-pulses will increase the population in the intermediate high energy state. The details for the derivation of the equation describing the probability amplitude for the population in the intermediate state can be found in Appendix \ref{statepopdynamics}. 
We find
\begin{widetext}
\begin{eqnarray}
&&\widetilde{c}_{i}    =\left[  \frac{i\Omega_{1}^{\ast}}{i\Delta-\frac{\gamma_{l}}{2}}e^{i(\mathbf{k}_{1}\cdot\mathbf{x}_{g})}\widetilde{c}_{g}+\frac
{i\Omega_{2}^{\ast}}{i(\Delta+\delta)-\frac{\gamma_{l}}{2}}e^{i(\mathbf{k}
_{2}\cdot\mathbf{x}_{e})}\widetilde{c}_{e}\right]  e^{-\frac{\gamma_{l}}{2}%
t} \nonumber\\
&&\hspace{7cm} -\left[  \frac{i\Omega_{1}^{\ast}}{i\Delta-\frac{\gamma_{l}}{2}}e^{i(\mathbf{k}%
_{1}\cdot\mathbf{x}_{g}+\Delta t)}\widetilde{c}_{g}+\frac{i\Omega_{2}^{\ast}%
}{i(\Delta+\delta)-\frac{\gamma_{l}}{2}}e^{i(\mathbf{k}_{2}\cdot\mathbf{x}_{e}+(\Delta+\delta)t)}\widetilde{c}_{e}\right]\nonumber\\ \label{cipopFinalv1}
\end{eqnarray}
\end{widetext}
Here $\Delta$ and $\delta$ are the single and two photon detunings as shown in Fig. {\ref{fig2_EnergyDiagram}}. As the pulse area increases due to the multiple $\pi$-pulses, this assumption breaks down. Here, we also solve the set of differential equations by considering the adiabatic elimination of the intermediate high energy state, but include the effects of spontaneous emission on the intermediate state dynamics we get the following dynamics for the ground and excited states

\begin{widetext}
\begin{subequations}
\begin{eqnarray}
\overset{\cdot}{\widetilde{c}}_{g}  &  = &\left[  \frac{\left\vert \Omega_{1}\right\vert^{2}}{i\Delta-\frac{\gamma_{l}}{2}}e^{i\Delta t}\widetilde{c}_{g}+\frac{\Omega_{1}\Omega_{2}^{\ast}}{i(\Delta+\delta)-\frac{\gamma_{l}}{2}}e^{i(\Delta_{kx}-\Delta t)}\widetilde{c}_{e}\right]e^{-\frac{\gamma_{l}}{2}t}
-\left[  \frac{\left\vert \Omega_{1}\right\vert ^{2}}{i\Delta-\frac{\gamma_{l}}{2}}\widetilde{c}_{g}+\frac{\Omega_{1}\Omega_{2}^{\ast}}{i(\Delta+\delta)-\frac{\gamma_{l}}{2}}e^{i(\Delta_{kx}-\delta t)}\widetilde{c}_{e}\right]
,\label{NewcgDynamics}\\
\overset{\cdot}{\widetilde{c}}_{e}  &  = &\left[  \frac{\left\vert \Omega_{2}\right\vert
^{2}}{i(\Delta+\delta)-\frac{\gamma_{l}}{2}}e^{i(\Delta+\delta)t}\widetilde{c}_{e}+\frac{\Omega_{2}\Omega_{1}^{\ast}}{i\Delta-\frac{\gamma_{l}}{2}}e^{i(-\Delta_{kx}-(\Delta+\delta)t)}\widetilde{c}_{g}\right] e^{-\frac{\gamma_{l}}{2}t}\nonumber\\
&&\hspace{6.0cm}-\left[  \frac{\left\vert \Omega_{2}\right\vert ^{2}}{i(\Delta+\delta)-\frac{\gamma_{l}}{2}}\widetilde{c}_{e}+\frac{\Omega_{2}\Omega_{1}^{\ast}%
}{i\Delta-\frac{\gamma_{l}}{2}}e^{i(-\Delta_{kx}-\delta t)}%
\widetilde{c}_{g}\right]  .\label{NewceDynamics}
\end{eqnarray}
\end{subequations}
\end{widetext}
Here $\Delta_{kx}=\mathbf{k}_{2}\cdot\mathbf{x}_{e}-\mathbf{k}_{1}%
\cdot\mathbf{x}_{g}$. Eqn. (\ref{NewceDynamics}) shows that the dynamics of the excited state is connected to the loss through $\gamma_{l}$. The terms within the first pair of square brackets  in Eqns. (\ref{NewcgDynamics}) and (\ref{NewceDynamics}) is
the AC term of the intermediate state which  oscillates and  decays. The second term is the
DC (stationary state) term for long enough time scales that the rapid oscillations
have dissipated. One can also see for the DC term that if $\gamma_{l}=0$ we return to
the closed quantum system atom interferometer dynamics when considering adiabatic elimination
of the intermediate state without spontaneous decay. For the full derivation of the state dynamics see Appendix \ref{statepopdynamics}. 

To demonstrate that this approach works for our model, we also performed a full density matrix approach with spontaneous decay into the loss state as well as the ground and excited states (see Appendix \ref{densmatrixdyn}). The full density matrix method compared to our loss state dominating the decay give the same results allowing us to continue with this approach.

\subsection{Error in the acceleration \label{sec_error}}

Now that we have a way to calculate the quantum information loss, we need to
understand how it will affect the acceleration measurement. While a co-sensor will approximately measure the total acceleration (beyond the maximum acceleration that the atom interferometer can measure: $2 \pi \times 1/(T^2 |\mathbf{k}_\mathrm{eff}|) $, see below), the atom interferometer measures the small deviation of the acceleration accurately. The phase-shift and the acceleration value deviation relate as follows \cite{JBF03,AVR15,BVE16,CF17},%
\begin{equation}
\mathrm{dev}(a) =\frac
{1}{T^{2}\left\vert \mathbf{k}_\mathrm{eff}\right\vert \cos\theta}\Delta(\Phi),
\label{netaccel}
\end{equation}
where $\mathrm{dev}(a)$ is the deviation of the acceleration, $T$ is the free evolution time of the atomic cloud, $\left\vert \mathbf{k}_\mathrm{eff}\right\vert $ is the effective wave
number, and $\theta$ is the angle between $\left\vert \mathbf{k}%
_\mathrm{eff}\right\vert $ and the net acceleration. The phase shift
of the interferometer is $\Delta(\Phi)$ and links directly with the population of the ground and
excited states such that $\Delta(\Phi)=\tan^{-1}\left(\left\vert c_{e}%
(t_{f})/c_{g}(t_{f})\right\vert \right)  $ where $t_{f}$ is the time at the
end of the interferometer pulse sequence. Increasing the number of Raman pulses from the usual $\pi/2-\pi-\pi/2$ pulse sequence will have our relationship be as follows \cite{MSK00,PB97}%
\begin{widetext}
\begin{align}
\mathrm{dev}(a) &=\left(\frac{1}{(2N_{R}T^{2}\left\vert \mathbf{k}_\mathrm{eff}\right\vert
-2(N_{R}+1)\left\vert \mathbf{k}_\mathrm{eff}\right\vert T\tau_{d})\cos\theta}\right) \Delta(\Phi) \nonumber \\
&= \alpha\tan^{-1}\left(  \sqrt{\left\vert \frac{c_{e}(t_{f})}{c_{g}(t_{f})}\right\vert ^{2}}\right),
\end{align}
\end{widetext}
where $\alpha=1/(2N_{R}T^{2}\left\vert \mathbf{k}_\mathrm{eff}\right\vert
-2(N_{R}+1)\left\vert \mathbf{k}_\mathrm{eff}\right\vert T\tau_{d})\cos\theta$ and $\tau_d$ is the time between sequential pulses. Here, $N_{R}=$ number of central $\pi$-Raman pulses as illustrated in Fig \ref{fig1_LMTmodel}. From Eqs. (\ref{NewcgDynamics}) and (\ref{NewceDynamics}), the acceleration will affect our dynamics through the $\Delta_{kx}$ term. We
consider the square of $\left\vert c_{e}(t_{f})/c_{g}(t_{f})\right\vert $ as
we can change the ratio in the arctangent to relate to the amount of quantum
information loss%
\begin{equation}
\left\vert \frac{c_{e}(t_{f})}{c_{g}(t_{f})}\right\vert^{2}=\frac{\left\vert
c_{e}(t_{f})\right\vert ^{2}}{ 1-\left\vert c_{e}(t_{f})\right\vert
^{2}-Q_{\mathrm{tot}}},
\end{equation}
where $Q_{\mathrm{tot}}=N_{R}Q + (N_{R}-1)Q + 2(Q/2)=mQ$ with $m=2N_{R}$ from the increase in $\pi$-pulses for LMT. The $Q/2$ term comes from the impact on the quantum information loss from the $\pi/2$ pulses. This relation can be made due to the symmetry of the Raman pulse sequence. From here we can look at the error in the acceleration measurement through the variance (i.e. noise) of the acceleration%
\begin{widetext}
\begin{equation}
 \mathrm{var}(\mathrm{dev}(a))=\alpha^{2}\left\langle\tan^{-1}\left(
\sqrt{\frac{\left\vert c_{e}(t_{f})\right\vert ^{2}}{1-\left\vert
c_{e}(t_{f})\right\vert ^{2}-Q_{\mathrm{tot}}}}\right) ^{2}\right\rangle.
\label{AccelVariance}
\end{equation}
\end{widetext}

As can be seen, we have to consider the variance of a ratio within an
arctangent function. To proceed, we will consider the variance in the population of
the excited state, $\left\vert c_{e}(t_{f})\right\vert ^{2}$. We measure the excited state population through photon counting giving a Poissonian distribution. Remembering from eqn. (\ref{NewceDynamics}), the excited state population will be correlated to the information loss through $\gamma_{l}$.

We will treat $Q_{\mathrm{tot}}$ as a random variable as well because spontaneous emission happens randomly. This will add yet another layer into the calculation of the variance as we have two variances that are correlated since the random spontaneous emission will affect both the quantum information loss and the final excited state population.

To continue, let us consider two stochastic variables $X$ and $Y$ with
expectation values $\left\langle X\right\rangle $ and $\left\langle
Y\right\rangle $ and the average of their ratio denoted by $R_{1}$. To find
the expectation value of the ratio, one can do a series expansion of $X/Y$
around $\left\langle X\right\rangle $ and $\left\langle Y\right\rangle $
\cite{VKVV00,MKAS77} to get
\begin{equation}
\left\langle \frac{X}{Y}\right\rangle =\frac{\left\langle X\right\rangle
}{\left\langle Y\right\rangle }+\mathrm{var}(Y)\frac{\left\langle X\right\rangle
}{\left\langle Y\right\rangle ^{3}}-\frac{\mathrm{cov}(X,Y)}{\left\langle
Y\right\rangle ^{2}}, \label{ratioexpect}
\end{equation}
where $\mathrm{var}(Y)$ is the variance of variable $Y$ and $\mathrm{cov}(X,Y)$ is the covariance
between $X$ and $Y$. The variables $X=\left\vert
c_{e}(t_{f})\right\vert ^{2}$ and $Y=1-\left\vert
c_{e}(t_{f})\right\vert ^{2}-Q_{\mathrm{tot}}$. We then can define $R_{1}=\left\vert c_{e}(t_{f})\right\vert ^{2}/1-\left\vert c_{e}(t_{f})\right\vert ^{2}-Q_{\mathrm{tot}}$. With the expectation value of the ratio between the two stochastic variables, the variance of the ratio can be determined
\begin{widetext}
\begin{align}
\mathrm{var}(R_{1})&=\left\langle \left(  R_{1}-\left\langle R_{1}\right\rangle\right)^{2} 
\right\rangle =\left\langle \left( \frac{\left\vert c_{e}(t_{f})\right\vert ^{2}}{1-\left\vert c_{e}(t_{f})\right\vert ^{2}-Q_{\mathrm{tot}}} -\left\langle\frac{ \left\vert c_{e}(t_{f})\right\vert ^{2}}{ 1-\left\vert c_{e}(t_{f})\right\vert ^{2}-Q_{\mathrm{tot}} }\right\rangle\right)^{2} \right\rangle, \nonumber \\ 
\mathrm{var}(R_{1})&=\frac{1}{n}\left(  \frac{\mathrm{var}(\left\vert c_{e}(t_{f})\right\vert ^{2})}{\left\langle 1-\left\vert c_{e}(t_{f})\right\vert ^{2}-Q_{\mathrm{tot}}\right\rangle ^{2}}+\frac{\left\langle \left\vert c_{e}(t_{f})\right\vert ^{2}\right\rangle ^{2}\mathrm{var}(1-\left\vert c_{e}(t_{f})\right\vert ^{2}-Q_{\mathrm{tot}})}{\left\langle 1-\left\vert c_{e}(t_{f})\right\vert ^{2}-Q_{\mathrm{tot}}\right\rangle
^{4}}-2\frac{\left\langle \left\vert c_{e}(t_{f})\right\vert ^{2}\right\rangle \mathrm{cov}(\left\vert c_{e}(t_{f})\right\vert ^{2},Q_{\mathrm{tot}})}{\left\langle 1-\left\vert c_{e}(t_{f})\right\vert ^{2}-Q_{\mathrm{tot}}\right\rangle
^{3}}\right) , \label{ratiovar}
\end{align}
\end{widetext}
where $n$ is the number of samples taken. We also need to consider that the denominator of the ratio depends on the difference between the excited state population and the quantum information loss. That being said, we will have two variables with correlated variances as they both depend on the number of Raman pulses. The adding or subtracting of the variance of these two correlated random variables is given by
\begin{widetext}
\begin{equation}
\mathrm{var}(\left\vert c_{e}(t_{f})\right\vert ^{2}) \pm \mathrm{var}(Q_{\mathrm{tot}}) = \mathrm{var}(\left\vert c_{e}(t_{f})\right\vert ^{2})+\mathrm{var}(Q_{\mathrm{tot}})\pm2\rho\sigma_{\left\vert c_{e}(t_{f})\right\vert ^{2}}\sigma_{Q_{\mathrm{tot}}}.
\end{equation}
\end{widetext}
Here $\sigma_{\left\vert c_{e}(t_{f})\right\vert ^{2}}$ and $\sigma_{Q_{\mathrm{tot}}}$ are the standard deviations, and $\rho$ relates to the covariance of the two variables as follows
\begin{equation}
\rho=\frac{\mathrm{cov}(\left\vert c_{e}(t_{f})\right\vert ^{2},Q_{\mathrm{tot}})}{\sqrt{\mathrm{var}(\left\vert c_{e}(t_{f})\right\vert ^{2})\mathrm{var}(Q_{\mathrm{tot}})}}.
\end{equation}

As seen above, the covariance between the excited state population and the quantum information loss will be needed. We determine this covariance by first rewriting the final excited state population and the quantum information loss due to their random error sources 
\begin{subequations}
\begin{align}
|c_{e} (t_f)|^2 &= \overline{|c_{e} (t_f)|^2}+\epsilon_{M}+\epsilon_{\mathrm{\Gamma}} , \label{excitedstatepop} \\
Q_{\mathrm{tot}} &= m\overline{Q} + \epsilon_{Q}.
\end{align}
\end{subequations}
Here $\epsilon_{M}$ and $\epsilon_{\mathrm{\Gamma}}$ are the random errors (having zero means) of the excited state population due to the measurement error and spontaneous emission, respectively. We note that $\epsilon_{M}$ is also the quantum efficiency of the detector. $\epsilon_{Q}$ is the random error in the quantum information loss due to the random spontaneous emission. Since the spontaneous emission will affect the final populations of the excited state, ground state, and loss state it can be seen that $\epsilon_{Q}$ will relate directly to $\epsilon_{\mathrm{\Gamma}}$ through $m\epsilon_{Q}=-2\epsilon_{\mathrm{\Gamma}} \gamma_{l} t_{\mathrm{tot}}$ (see Appendix \ref{appendixB}). The covariance between the final excited state population and the quantum information loss is defined by
\begin{align}
\mathrm{cov}(\left\vert c_{e}(t_{f})\right\vert ^{2}, Q_{\mathrm{tot}}) = \langle\left\vert c_{e}(t_{f})\right\vert ^{2} Q_{\mathrm{tot}} \rangle - \langle\left\vert c_{e}(t_{f})\right\vert ^{2}\rangle \langle Q_{\mathrm{tot}} \rangle . 
\end{align}
Due to the only correlating values being the errors due to the random spontaneous emission, we get the covariance to come out to be as follows (for full derivation see Appendix \ref{appendixB})
\begin{align}
\mathrm{cov}(\left\vert c_{e}(t_{f})\right\vert ^{2}, Q_{\mathrm{tot}}) = \langle\epsilon_{\mathrm{\Gamma}}\epsilon_{Q}\rangle = - \frac{1}{2\gamma_l t_\mathrm{tot}}\mathrm{var}(Q_\mathrm{tot}),
\end{align}
where $\mathrm{var}(Q_\mathrm{tot})$ is the variance in the quantum information loss, defined in Appendix \ref{appendixB}.

With the ability to determine the variance of the ratio, the next step is to
linearize the arctangent function. We can do this since we are interested in small deviations of the measured acceleration. This allows us to linearly approximate the arctangent function. This is done by taking the first-order expansion of the Taylor series:
\begin{equation}
\tan^{-1} (x+\delta x) \simeq \tan^{-1}(x)+\left(\frac{1}{1+x^{2}}\right)\delta x.
\end{equation}
Here $x$ is our reference point and $\delta x$ is a small deviation from this reference point. For our system, the value of $x$ will be the expectation value of the ratio $R_{1}$ (Eqn. (\ref{ratioexpect})) and $\delta x$ will be the standard deviation of $R_{1}$ (i.e. square root of Eqn. (\ref{ratiovar})). Let us now rewrite the linearized arctangent in terms of $R_{1}$,  
\begin{equation}
\tan^{-1} (R_{1}+\mathrm{dev}(R_{1})) \simeq \tan^{-1}(R_{1})+\left(\frac{\mathrm{dev}(R_{1})}{1+(R_{1})^{2}}\right).
\end{equation}
With the arctangent function linearized, we can now consider the variance in the phase shift which will relate to the linearized function as follows
\begin{equation}
\mathrm{var}(\phi) =\mathrm{var}(R_{1})
d^{2}+2db\sqrt{\mathrm{var}(R_{1})} +b^{2}.
\end{equation}
where $d=(1/(1+(R_{1})^{2}))$ is the slope and $b=\mathrm{tan}^{-1}(R_{1})$ is the y-intercept. Here we have related the variance of the phase shift, $\mathrm{var}(\phi)$, with the arctangent of the ratio explained in Eq. (\ref{AccelVariance}) which gives us the noise from the
spontaneous decay
\begin{equation}
\mathrm{var}(\mathrm{dev}(a)) = \alpha^{2}\left (\mathrm{var}(R_{1})d^{2} +2db\sqrt{\mathrm{var}(R_{1})} + b^{2} \right).
\end{equation}
It should be noted that this will relate to other noise
sources that affect $\phi$ but we are currently just concerned with the noise
from spontaneous decay of the intermediate high-energy state.

Now that we have a noise model due to spontaneous decay, we next want to determine a figure of merit for the error of the measured acceleration. This will relate more directly to the measurement taken in a given experiment. We start by looking at the deviation of the small acceleration changes that are measured. This is given by
\begin{equation}
\delta (\mathrm{dev}(a))=a_{\mathrm{tr}}-\mathrm{dev}(a) \label{meanvaldev}
\end{equation}
where $a_{\mathrm{tr}}$ is the true value of the small changes in acceleration trying to be measured and $\mathrm{dev}(a)$ is the small deviation in acceleration measured by the atom interferometer. We want our figure of merit to encompass the DC offset described by the mean value deviation (Eq. (\ref{meanvaldev})) as well as the AC random fluctuation described by the variance. We can do this by considering the following
\begin{widetext}
\begin{align}
\langle \delta (\mathrm{dev}(a))^{2}\rangle &=\langle\left(a_{\mathrm{tr}}-\mathrm{dev}(a)\right)^{2}\rangle \nonumber \\
&=a_{\mathrm{tr}}^{2}-2a_{\mathrm{tr}}\langle \mathrm{dev}(a)\rangle +\langle \mathrm{dev}(a)^{2}\rangle  \nonumber \\
&=a_{\mathrm{tr}}^{2} -2a_{\mathrm{tr}}\langle \mathrm{dev}(a)\rangle +\mathrm{var}(\mathrm{dev}(a)) +\langle \mathrm{dev}(a)\rangle ^{2} \nonumber \\
&= (a_\mathrm{tr} - \langle\mathrm{dev}(a) \rangle)^2 + \mathrm{var} (\mathrm{dev}(a)). \label{FOM}
\end{align}
\end{widetext}
Here, we used the fact that we can write the variance of $\mathrm{dev}(a)$ as $\mathrm{var}(\mathrm{dev}(a))=\langle (\mathrm{dev}(a))^{2}\rangle -\langle \mathrm{dev}(a)\rangle ^{2}$. This gives us our combined error term for both a static DC offset (the first term on the right-hand side) and the noise, AC random fluctuation (the second term on the right-hand side), which we consider a figure of merit for the performance of the acceleration sensor atom interferometer. It should be noted that the ideal case without errors and noise would have this figure of merit be zero.

\section{Results\label{sec_results}}

With the ability to calculate the variance and the error figure of merit ($\langle \delta (\mathrm{dev}(a))^{2}\rangle = \mathrm{FOM}$), we simulated the dynamics of the atom interferometer and performed a statistical analysis on the FOM. We considered a system of Rubidium-85 ($^{85}$Rb) atoms with system parameters described in Table \ref{table_1} which were taken from Steck \textit{et al} \cite{DAS21} to determine Rabi frequencies, the lengths of the $\pi-$ and $\pi/2-$ pulses, and the spontaneous decay rate. The single-photon detuning ($\Delta$) was set at $9$ $\mathrm{GHz}$ and the two-photon detuning ($\delta$) was set to zero as this could be done with fine-tuning of the lasers (e.g. $\omega_{1}-\omega_{2}=\omega_{ge}$). We assumed that a co-sensor is able to resolve the approximate acceleration and, thus, the location of fringe lobes in the atom interferometer measurements \cite{MGB09,CFT18}, while our atom interferometer measures small deviations in the acceleration. Lastly, we consider the beam and atomic cloud will be oriented in a way such that $\cos\theta=1$ which is a usual approximation made during experiment. We then simulate the LMT pulses by numerically solving the differential equations of the dynamics (Eqns. (\ref{NewcgDynamics}) and (\ref{NewceDynamics})) using a 4th order Runge-Kutta method. Multiple simulations with changing starting positions and velocities are run for each different number of $\pi$-Raman
pulses to get the statistical data needed to calculate the variance and the $\mathrm{FOM}$. The velocities change in such a manner so that the root-mean-square of the atom cloud is the MOT temperature.

\begin{table}[h]
\begin{center}%
\begin{tabular}
[c]{|l|l|}\hline
\multicolumn{2}{c}{\textbf{System Parameters}} \\
\hline
Free evolution time & $100$ $\mathrm{ms}$ \\\hline
Time between successive pulses & $150$ $\mathrm{\mu s}$ \\\hline
$\pi$-pulse length ($t_{\pi}$) & $2.00$ $\mathrm{\mu s}$ \\\hline
$\pi /2$ pulse length ($t_{\pi /2}$) & $1.00$ $\mathrm{\mu s}$ \\\hline
Single-photon detuning ($\Delta$) & $2 - 20$ $\mathrm{GHz}$ \\\hline
Two-photon detuning ($\delta$) & $0 - 63$ $\mathrm{kHz}$ \\\hline
Rabi frequency ($\Omega_{1}=\Omega_{2}$) & $212$ $\mathrm{MHz}$ \\\hline
Raman laser wavelength ($\lambda_{1}=\lambda_{2}$) & $780$ $\mathrm{nm}$ \\\hline
Spontaneous decay rate ($\Gamma$) & $38.117$ $\mathrm{MHz}$ \\\hline
Mass of Rb$^{85}$ & $1.419 \cdot 10^{-25}$ $\mathrm{kg}$ \\\hline
MOT temperature & $2$ $\mathrm{\mu K}$ \\\hline
$a_{\mathrm{tr}}$ & $1.85 \cdot 10^{-5}$ $\mathrm{m/s^{2}}$\\\hline
\end{tabular}
\end{center}
\caption{System parameters used in our numerical simulations for the LMT multi-Raman pulse atom interferometer acceleration sensor. The Rb$^{85}$ D-line data from Steck \textit{et al} \cite{DAS21} was used to determine these parameters for the simulation.}
\label{table_1}
\end{table}

\subsection{Numerical results of FOM depending on the number of Raman pulses\label{sec_PhotonCountResults}}

With our FOM determined, we can start to analyze the limitations of an LMT atom interferometer acceleration sensor. Thus far we have not directly taken into account uncertainty in the actual measurement process. Measuring the acceleration requires measuring the population of the excited state. This will bring in an error that is uncorrelated to the spontaneous decay but gives an extra piece of uncertainty in the measurement which we can add into our calculation. The measurement of the excited state is done by driving the excited state, $\left\vert e\right\rangle$, to a state not part of the Raman process and counting the photons which come out. Uncertainty in photon counting comes from systematic errors such as a branching process or the excited state decaying before the driving occurs. This means that the population measured may not be the actual population achieved from the atom interferometer pulse sequence. We thus consider that our excited state population will be given by Eq. (\ref{excitedstatepop}) and we can implement the error due to photon counting ($\epsilon_{M}$). 

\begin{figure}[ptb]
\centering
\includegraphics[width=\columnwidth]{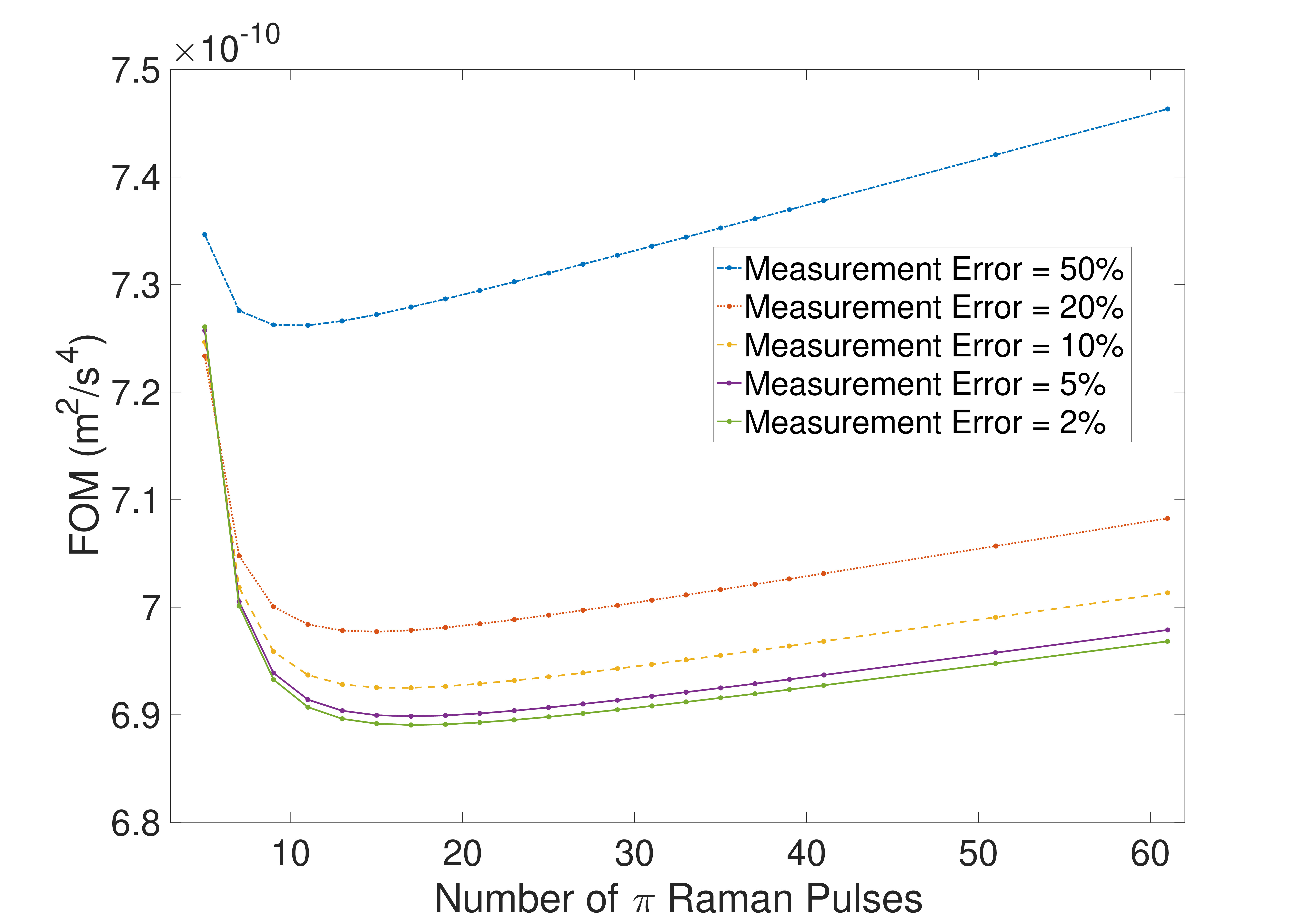}\caption{The error figure of merit of the measured acceleration for different measurement error percentages: 50\% (blue dashed-dotted), 20\% (red dashed), 10\% (yellow dashed), and 5\% (purple solid), and 2\% (green solid). The single-photon large detuning is set to $\Delta=9$ GHz. The dependence on $N_{R}^{-2}$ of the noise dominates for low pulse systems. A minimum in the error is reached for the system of $17$ $\pi$-Raman pulses. Then the quantum information loss from the undesired spontaneous decay starts to take over and increases the error in the measurement.}
\label{fig3_MeanValueDev}%
\end{figure}

Incorporating this change into our analysis for a $\Delta=9$ $\mathrm{GHz}$ system, looking at cases where the photon counting uncertainty ranges from $2\%$ to $50\%$, and considering $Q_{\mathrm{tot}}$ is random we calculate the $\mathrm{FOM}$ as a function of the number of $\pi$-Raman pulses as seen in Fig. \ref{fig3_MeanValueDev}. Here the data shows that for low pulse numbers (e.g. $<10$), the $\mathrm{FOM}$ is dominated by the $N_{R}^{-2}$ noise dependence. For example, $17$ $\pi$-Raman pulses (i.e. $34\hbar k_\mathrm{eff}$ least error momentum transfer) produces the minimum FOM for the cases considering a measurement uncertainty of $10\%$ or less. 

Table \ref{table_2} shows that there will be no change in pulse amount which gives a minimized $\mathrm{FOM}$ up to $10\%$ error. Once the error exceeds $10\%$, fewer pulses are possible before we reach a minimum FOM where the DC offset starts to dominate. Our data shows that the measurement uncertainty due to the quantum efficiency of the detector will impact the performance and could cause noticeable issues in measuring the acceleration if not considered.
\begin{table}[h]
\begin{center}%
\begin{tabular}
[c]{|c|c|}
\hline
\textbf{Detection Uncertainty ($\epsilon_{M}$)} & \textbf{Minimum-FOM $N_{R}$}%
\\\hline
$0.02$ & $17$\\\hline
$0.04$ & $17$\\\hline
$0.10$ & $17$\\\hline
$0.20$ & $15$\\\hline
$0.50$ & $11$\\\hline
\end{tabular}
\end{center}
\caption{The $\pi$-Raman pulse number which gave the minimum mean value deviation when considering different detection error percentages for the case where $\Delta=9$ $\mathrm{GHz}$. Up to $10\%$, there is no noticeable change. Exceeding $10\%$, the number of $\pi$-Raman pulses that can be used before spontaneous decay error starts to take over goes down.}%
\label{table_2}
\end{table}

\begin{figure}[ptb]
\centering
\includegraphics[width=\columnwidth]{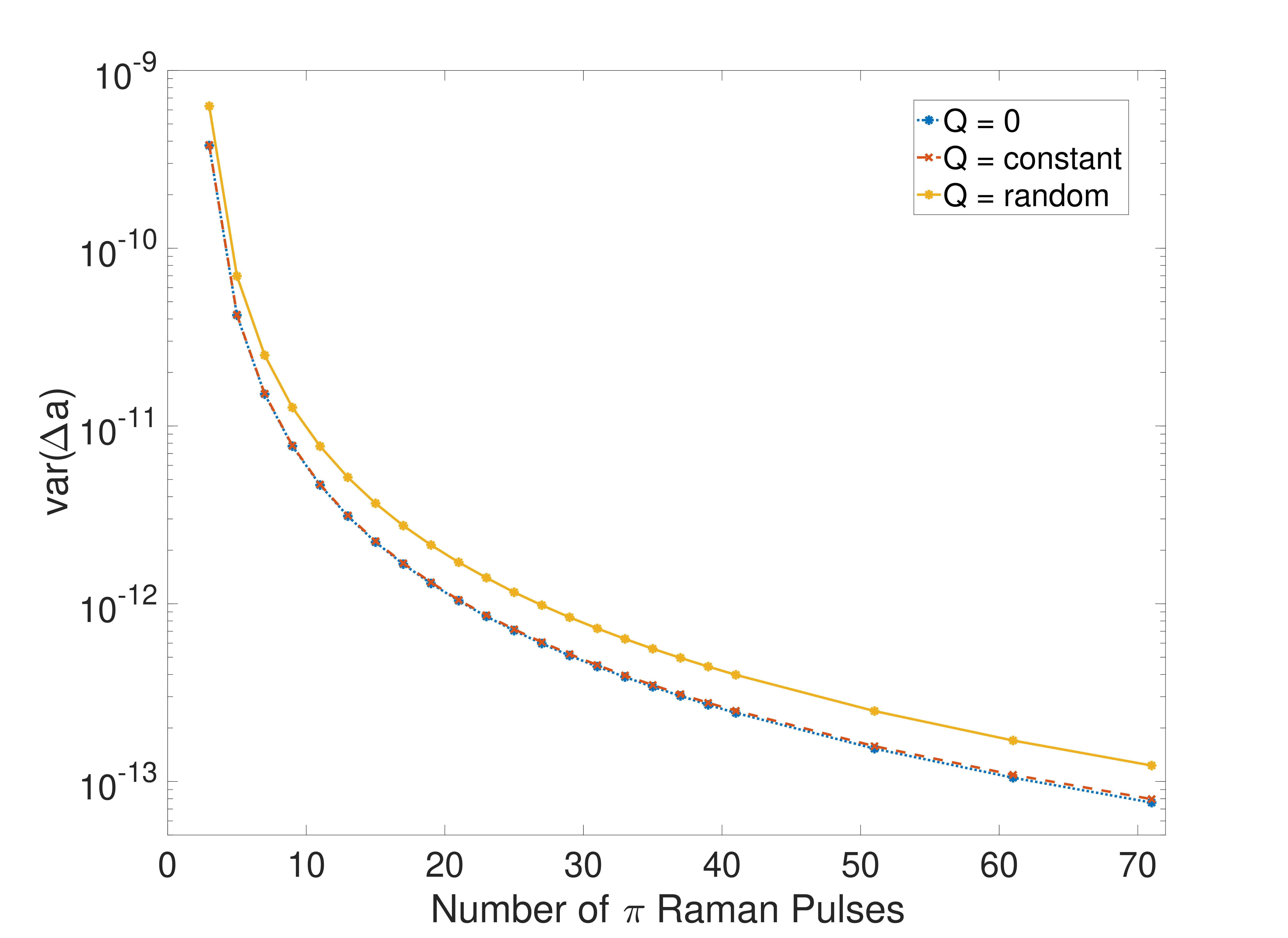}\caption{Variance in the acceleration measurement for the cases of $Q=$ random variable (solid orange), $Q=$ constant (dashed red), and $Q=0$ (dotted blue). For each case, the noise follows a dependence on the inverse square of the number of $\pi$-Raman pulses.}
\label{fig2_Noise}
\end{figure}

Seeing that our FOM trend changes over the increase in number of $\pi$-Raman pulses, we wanted to gain a better understanding of which term (AC or DC) was causing the change. We did this by looking at the effect of the number of Raman pulses on the variance of the acceleration measurement ($\mathrm{var}(\Delta a)$, i.e. the AC term). For this, we considered three different cases. First, we considered a $Q_{\mathrm{tot}}=0$ lossless (i.e., no spontaneous decay) case to gain intuition on the pure measurement-noise dependence of the number of Raman pulses. Next, we considered the quantum information loss being a discrete value, which is approximately valid for a case of an atom cloud with a large number of atoms typically used for atom interferometry. Lastly, we considered the quantum information loss to be a random variable, simulating the probabilistic spontaneous decay along with the excited state population, in the variance calculation as described above (section (\ref{sec_error})). Figure \ref{fig2_Noise} shows the variance for each of the three cases as a function of the number of Raman pulses.

The data shows that for each case the variance follows a dependence of $N_{R}^{-2}$. This is expected for the case of $Q=0$ as the dependence on the number of pulses will be solely in the constant $\alpha^{2}$ defined earlier which is directly proportional to $N_{R}^{-2}$. For the other cases, this trend is largely due to the fact that the amount of quantum information loss does not affect the overall variance much due to the low probability of spontaneous decay at low pulse numbers. The probability of spontaneous decay increases with increasing pulse number, but never to a point where it overtakes the $N_{R}^{-2}$ scaling of the pulse number in the noise.
We also see that for the case of $Q_{\mathrm{tot}}=0$ and $Q_{\mathrm{tot}}$ being a constant value there is little difference in the noise profiles. This is due to the fact that for these two cases $\mathrm{var}(Q_{\mathrm{tot}})=0$ and $\mathrm{cov}(\left\vert c_{e}(t_{f})\right\vert ^{2}, Q_{\mathrm{tot}})=0$. This will give new forms for Eqn. (\ref{ratiovar}) 
\begin{widetext}
\begin{align}
\mathrm{var}(R_{1})_{Q_{\mathrm{tot}}=0}&=\frac{1}{n}\left(  \frac{\mathrm{var}(\left\vert c_{e}(t_{f})\right\vert ^{2})}{\left\langle 1-\left\vert c_{e}(t_{f})\right\vert ^{2}\right\rangle ^{2}}+\frac{\left\langle \left\vert c_{e}(t_{f})\right\vert ^{2}\right\rangle ^{2}\mathrm{var}(1-\left\vert c_{e}(t_{f})\right\vert ^{2})}{\left\langle 1-\left\vert c_{e}(t_{f})\right\vert ^{2}\right\rangle
^{4}}\right) ,\nonumber\\
\mathrm{var}(R_{1})_{Q_{\mathrm{tot}}=\mathrm{cons}}&=\frac{1}{n}\left(  \frac{\mathrm{var}(\left\vert c_{e}(t_{f})\right\vert ^{2})}{\left\langle 1-\left\vert c_{e}(t_{f})\right\vert ^{2}-Q_{\mathrm{tot}}\right\rangle ^{2}}+\frac{\left\langle \left\vert c_{e}(t_{f})\right\vert ^{2}\right\rangle ^{2}\mathrm{var}(1-\left\vert c_{e}(t_{f})\right\vert ^{2})}{\left\langle 1-\left\vert c_{e}(t_{f})\right\vert ^{2}-Q_{\mathrm{tot}}\right\rangle
^{4}}\right) .
\end{align}
\end{widetext}
Now it should be noted that $\left\vert c_{g}(t_{f})\right\vert ^{2}=1-\left\vert c_{e}(t_{f})\right\vert ^{2}$ for the $Q_{\mathrm{tot}}=0$ case and $\left\vert c_{g}(t_{f})\right\vert ^{2}=1-\left\vert c_{e}(t_{f})\right\vert ^{2}-Q_{\mathrm{tot}}$ for the $Q_{\mathrm{tot}}$ being a constant case. With constant $Q_{\mathrm{tot}}$, any increase of $Q_{\mathrm{tot}}$ will result in $\left\vert c_{e}(t_{f})\right\vert ^{2}$ decreasing by the same amount. This will have our ratio in the arctangent affected minimally by a constant quantum information loss. When the covariance is incorporated, we see a slight increase in the noise due to having a second randomly fluctuating variable.

Looking back at Fig. \ref{fig3_MeanValueDev}, we see that initially the FOM curve is dominated by the AC term described by Fig. \ref{fig2_Noise}. As the number of pulses continues to increase, the FOM becomes dominated by the DC term. This is due to a few different factors. The DC error term in the FOM is
\begin{widetext}
\begin{align}
 &( a_\mathrm{tr} - \langle \mathrm{dev}(a)\rangle)^2 = \alpha^2 \left( \tan^{-1} \left|\frac{\tilde{c}_e (t_f)}{\tilde{c}_g (t_f)} \right| - \tan^{-1} \left|\frac{c_e (t_f)}{c_g (t_f)} \right| \right)^{2},
\end{align}
\end{widetext}
where $|\tilde{c}_{e,g} (t_f)|^2$ are the ideal populations without information loss (i.e., $Q_\mathrm{tot} = 0$). For nonzero $Q_\mathrm{tot}$ the populations shift to $|c_{e,g} (t_f)|^2=1-Q-|c_{g,e} (t_f)|^{2}$. It is difficult to see directly from the analytical relation how the nonzero $Q_\mathrm{tot}$ starts to increase the DC error. To explain the increase in DC error, let us consider again how the number of $\pi$ pulses affects the atom interferometer. 

As the number of $\pi$ pulses increases, the atom interferometer becomes more sensitive to the acceleration present due to the increase in momentum separation of the atomic states. This brings asymmetry in the ground and excited state wavefunctions which will bring error into the measurement. It has been seen in experiment that the efficiency of the $\pi$ pulses begins to diminish with increasing pulse numbers \cite{MSK00,BKK13,KBK15}. This could be why experiments have used composite pulses over standard Raman pulses to realize a multi-Raman pulse LMT atom interferometer \cite{BKK13}. Adding in another error source such as loss of quantum information, the efficiency of the $\pi$ pulses will get even worse. Even though the analytical intuition does not give an initial insight into quantum information loss increasing the DC error, our numerical simulations are showing the opposite (i.e. an increase in $Q_\mathrm{tot}$ leads to an increase in DC error).

\subsection{Effects of single-photon detuning \label{sec_DeltaResults}}

\begin{figure}[ptb]
	\centering
	\includegraphics[width=\columnwidth]{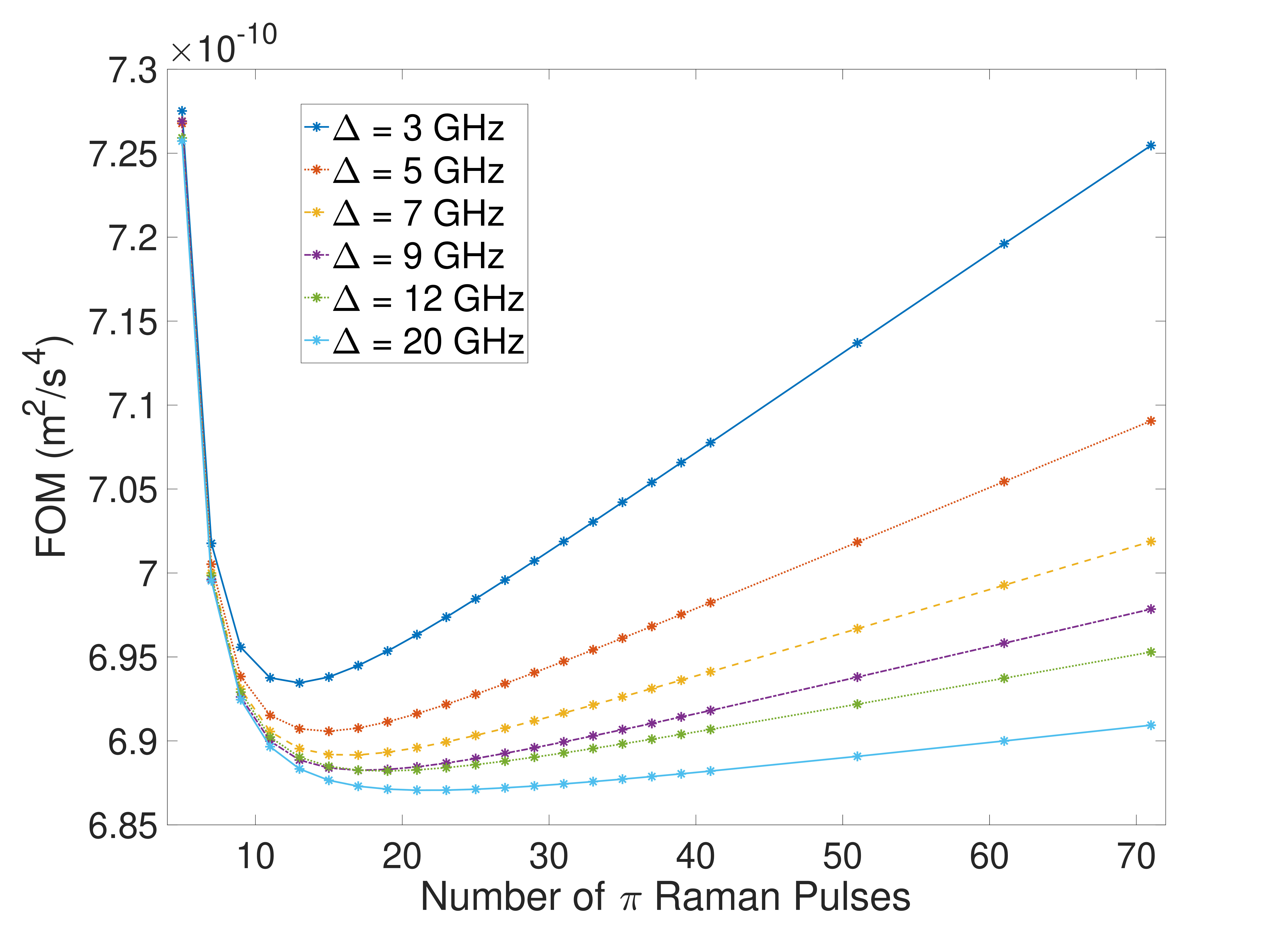}\caption{The error figure of merit of the measured acceleration for the cases of $\Delta=20$ $\mathrm{GHz}$ (solid light-blue), $\Delta=12$ $\mathrm{GHz}$ (dotted green), $\Delta=9$ $\mathrm{GHz}$ (dashed-dotted purple), $\Delta=7$ $\mathrm{GHz}$ (dashed orange), $\Delta=5$ $\mathrm{GHz}$ (dotted red), and $\Delta=3$ $\mathrm{GHz}$ (solid dark blue). The measurement error is set to $2\%$. As the value for $\Delta$ decreases, the minimum error happens for smaller pulse amounts. The steepness in the increase in error also begins to rise with lower $\Delta$ due to the higher probability of spontaneous decay.}%
	\label{fig4_DiffDeltaMVD}%
\end{figure}

Thus far, we have considered a large single-photon detuning of $\Delta=9$ $\mathrm{GHz}$, but certain experimental setups may require lower or higher detuning amounts. Recent experiments have used detunings on the order of $2$ to $4$ $\mathrm{GHz}$ \cite{MSK00,BKK13,KBK15}. Figure \ref{fig4_DiffDeltaMVD} shows the $\mathrm{FOM}$ as a function of the number of $\pi$-Raman pulses for detunings ranging from $3$ to $20$ $\mathrm{GHz}$. The data is showing that as we decrease our detuning, the number of pulses to get our minimum FOM is getting smaller along with the FOM starting to get steeper with an increased number of pulses. For detunings of $7$ and $9$ $\mathrm{GHz}$ the $\mathrm{FOM}$ minimizes at $17$ pulses. This would give a small-error LMT of $34\hbar$$k_\mathrm{eff}$ before spontaneous decay starts to take over. For detunings of $12$ and $20$ $\mathrm{GHz}$, the $\mathrm{FOM}$ minimized at $19$ and $21$ pulses respectively giving a small-error LMT of $38$ and $42\hbar$$k_\mathrm{eff}$ before spontaneous decay takes over.

\begin{figure}[ptb]
	\centering
	\includegraphics[width=\columnwidth]{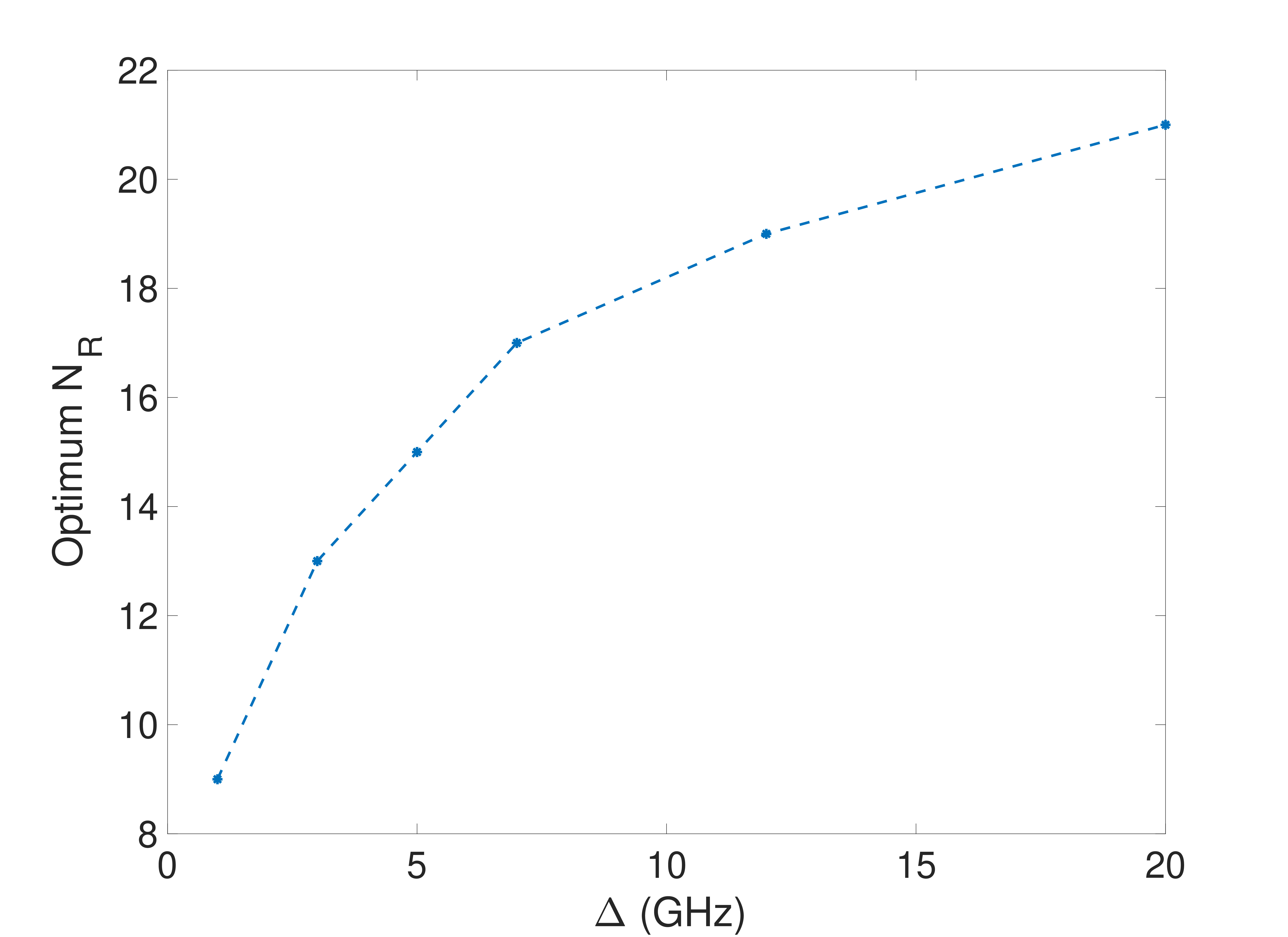}\caption{The $\pi$-Raman pulse number which gave the optimal error figure of merit when considering different large detuning ($\Delta$) values. As $\Delta$ goes down the number of $\pi$-Raman pulses that can be used before spontaneous decay starts to overtake the $N_{R}^{-2}$ noise dependence of the error goes down.}
	\label{fig5_DeltaVsMaxPulse}
\end{figure}

Figure \ref{fig5_DeltaVsMaxPulse} shows the pulse amount which gave the minimum $\mathrm{FOM}$ for all the different detuning values presented in Fig. \ref{fig4_DiffDeltaMVD}, plus the case for $\Delta = 1$ $\mathrm{GHz}$ which is not shown. The $1$ $\mathrm{GHz}$ case has the $\mathrm{FOM}$ minimum occur at $9$ $\pi$-Raman pulses. The $\Delta=1$ $\mathrm{GHz}$ case also shows such a steep increase after hitting its minimum that plotting it with others would drown out any noticeable data from the graphs. Continuing to decrease $\Delta$ will lead to an increase in spontaneous decay. This is the big issue with STIRAP as it relies on having $\Delta\approx0$ \cite{SBK92}. The data shows here that the detuning parameter will need to be finely tuned as too low a detuning will cause spontaneous decay to quickly take over the $N_{R}^{-2}$ dependence. High detunings will allow for larger and larger LMT but will require higher laser intensity. If an experiment wants to increase the detuning from $1$ $\mathrm{GHz}$ to $10$ $\mathrm{GHz}$ this would require increasing the intensity by that same factor of $10$. This increase in beam intensity will degrade the overall contrast of the atom interferometer \cite{CSP09}.

\subsection{Effects of two-photon detuning \label{sec_deltaResults}}

Thus far, we have considered the Raman two-photon detuning, $\delta$, to be zero as this can in principle be fulfilled with fine-tuning. This fine-tuning is not trivial and there could be a case where this value will still be $>0$ after fine-tuning or even desired (such as for frequency swept ARP \cite{KBK15}). The value for $\delta$ is defined by the following \cite{BKK13},
\begin{equation}
\delta = \left( \omega_{1} - \omega_{2}\right) - \left(\omega_{\mathrm{HFS}} - \textbf{k}_{\mathrm{eff}}\cdot\textbf{v} + \frac{\hbar\textbf{k}_{\mathrm{eff}}^{2}}{2m}\right) + \delta_{AC} . \label{deltadefinition}
\end{equation}
Here $\omega_{1,2}$ are the Raman laser frequencies, $\omega_{\mathrm{HFS}}$ is the hyperfine splitting, $\textbf{k}_\mathrm{eff}\cdot\textbf{v}$ constitutes the Doppler shift, $\hbar\textbf{k}_\mathrm{eff}^{2}/2m$ is for recoil shifts, and $\delta_{AC}$ is the AC Stark shifts. We have considered $\delta=0$ as we have set $\omega_{1} = \omega_{2}$. As seen in Eqn. (\ref{deltadefinition}) there are other factors that need to be mitigated and thus far we have considered fine-tuning of the experiment would have these effects go to zero.

Since fine-tuning in a way that will have $\delta=0$ is non-trivial, it is worth looking into the impact of a non-zero $\delta$. To study systems that would take into account a non-zero $\delta$, we consider the experiment for LMT by Butts \textit{et al} which locked $\Delta=3.5$ $\mathrm{GHz}$ and had $\delta=52$ $\mathrm{kHz}$ \cite{BKK13}. We also consider the work of McGuirk \textit{et al} which locked $\Delta=2$ $\mathrm{GHz}$ \cite{MSK00}. The work of McGuirk \textit{et al} does not quote a value for $\delta$. As they were working with Cesium-133 ($^{133}\mathrm{Cs}$), we used the D-line data from Steck \textit{et al} \cite{DAS19} to consider recoil shifts and Doppler shifts to get a plausible value for $\delta$ in their experiment if these effects were not mitigated. We last looked at cases for $\Delta=9$ $\mathrm{GHz}$ with a $\delta$ that has shifted due to Doppler and recoil shifts for the $^{85}$Rb system we are interested in and the case where fine-tuning has brought $\delta=0$. We also consider a measurement uncertainty of $2$.
\begin{figure}[ptb]
	\centering
	\includegraphics[width=\columnwidth]{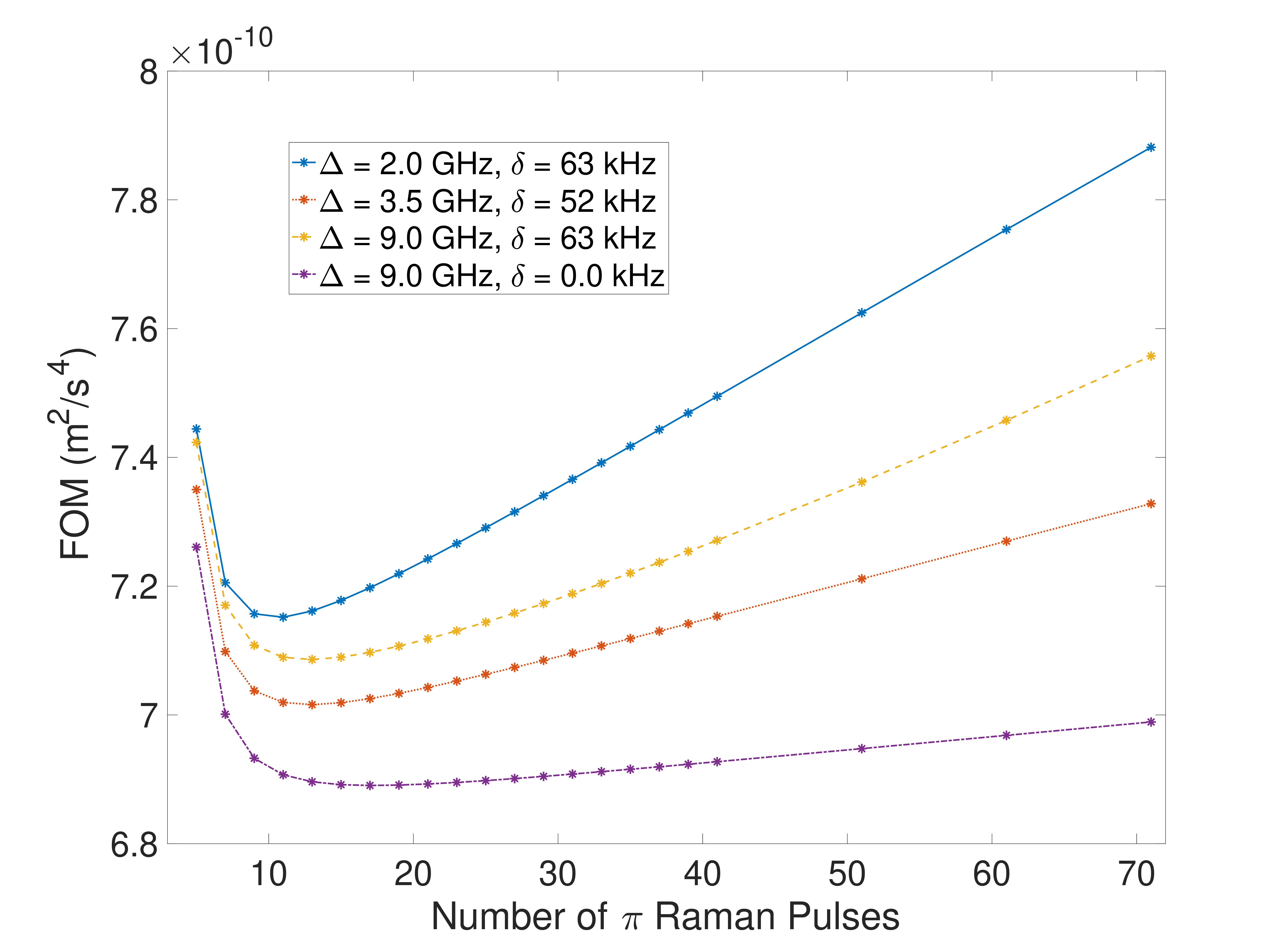}\caption{The error figure of merit for the measured acceleration when considering a non-zero $\delta$ detuning. The measurement error is set to $2\%$. We consider the $\Delta=9$ $\mathrm{GHz}$ case for $\delta=0$ (purple dashed-dotted) and $\delta=63$ $\mathrm{kHz}$ (orange dashed) which is possible when considering the $\delta$ shift due to photon recoil and Doppler shift of an atom at the recoil velocity in $^{85}$Rb \cite{DAS21}. The other cases consider the experimental $\Delta$ and $\delta$ used by Butts \textit{et al} \cite{BKK13} (red dotted) and McGuirk \textit{et al} with an assumed possible shift of $\delta$ as none was given \cite{MSK00} (blue solid). We see that the $\delta$ detuning plays a big role in reducing the performance as even increasing the $\Delta$ detuning with the same $\delta$ saw no improvement in the minimum error.}%
	\label{fig6_DiffdeltaError}%
\end{figure}

Figure \ref{fig6_DiffdeltaError} shows the $\mathrm{FOM}$ as a function of the number of $\pi$-Raman pulses for each of the cases. We see similar to Fig. \ref{fig4_DiffDeltaMVD} that a larger $\Delta$ detuning is better, but what is more apparent is that when we consider similar $\delta$ the increase in $\Delta$ shows less increase in performance. For instance, the $\Delta = 9$ $\mathrm{GHz}$ and $\Delta=2$ $\mathrm{GHz}$ cases show a minimum $\mathrm{FOM}$ for the $13$ and $11$ $\pi$-Raman pulses respectively when $\delta=63$ $\mathrm{kHz}$. Whereas the $\Delta=9$ $\mathrm{GHz}$ case with $\delta=0$ gives a minimum $\mathrm{FOM}$ for $17$ $\pi$-Raman pulses. The $\Delta=3.5$ $\mathrm{GHz}$ case with a slightly smaller $\delta$ even shows the same $\mathrm{FOM}$ minimum pulse number as the $\Delta=9$ $\mathrm{GHz}$ case. We even see a steeper increase in error after the minimum for the case of non-zero $\delta$. Even with a large increase in the detuning $\Delta$ (e.g. $2$ to $9$ $\mathrm{GHz}$), a moderate shift in the $10$ $\mathrm{kHz}$ range (which is plausible) for $\delta$ will bring a noticeable loss of performance that will need to be compensated for through fine-tuning.

When looking at the system similar to Butts \textit{et al}, we see for the $6\hbar k_\mathrm{eff}$ and $10\hbar k_\mathrm{eff}$ an average quantum information loss due to spontaneous decay per pulse of $0.26\%$. This is in good agreement with the estimate of $0.25\%$ they give in their work for the same pulse amounts \cite{BKK13}. Though again it should be noted we consider a system of $^{85}$Rb atoms and their experiment used $^{133}$Cs which will have different system parameters. As for the work done by McGuirk \textit{et al}, they estimate a per pulse spontaneous decay of $1\%$ \cite{MSK00}. When looking at our case that estimates a plausible $\delta$ value, we are seeing a per pulse quantum information loss from spontaneous decay of $0.5\%$. The difference could be explained as we are considering a lower value for $\delta$ than they had for their experiment as well as the difference in the atomic system (again $^{85}$Rb compared to $^{133}$Cs). When Eqns. (\ref{NewcgDynamics}) and (\ref{NewceDynamics}) are considered, the spontaneous decay highly affects the AC term both in the amplitude and in the phase, whereas it only plays a role in the amplitude for the DC term. However, the detuning $\delta$ will be a factor in the phase for both terms. In fact, since $\Delta\gg\delta,\gamma$, the detuning $\delta$ has a more noticeable effect on the state populations of the ground and excited states since it solely affects the phase of the DC term. This means a difference in our estimate for the spontaneous decay and that of McGuirk \textit{et al} could be that they had an increase in the ground state population from an appreciable $\delta$ detuning causing decay of the excited state giving an overestimate of the actual spontaneous decay amount. In summary, the large detuning $\Delta$ will affect the overall amount of spontaneous decay and the detuning $\delta$ will affect the efficiency of the Raman pulses populating the excited state.

\section{Conclusions and Discussions}

The work presented here has investigated the limitations of a LMT multi-Raman pulse atom interferometer acceleration sensor when considering undesired spontaneous decay.  We began by considering an open quantum system where an increase in optical interactions gives rise to an increased chance of spontaneous decay. We modeled the open quantum system by considering the Markovian nature of this decay and applied the Lindblad master equation in the Schr\"{o}dinger picture. We were then able to formulate the new atomic state dynamics taking into account the undesired spontaneous decay of the intermediate high energy state. We numerically simulated the atom interferometer acceleration sensor using the Runge-Kutta 4th order integration for a system of $^{85}$Rb atoms using parameters from Steck \textit{et al} \cite{DAS21}. We then derived the variance and combined it with the mean-value deviation to determine an error figure of merit (FOM) of the measured acceleration.

We found that the error FOM had a minimum at a certain number of $\pi$-Raman pulses ($N_R$). This minimum was explainable as follows. For low number of $\pi$-Raman pulses, the quantum information loss ($Q_\mathrm{tot}$) is small and, hence, the FOM was dominated by $N_{R}^{-2}$ dependent error reduction. For large number of $\pi$-Raman pulses, the FOM was dominated by substantial quantum information loss, which increased the DC offset error. These two competing trends produced the optimal number of $\pi$-Raman pulses that accomplishes the overall minimum error. In our numerical analysis, we also incorporated that the measurement of the excited state population can have uncertainty uncorrelated with the quantum information loss from undesired spontaneous decay. We incorporated this into our error analysis as a percentage of the excited state population lost during measurement. As the measurement uncertainty increased, the LMT with minimum error began to decrease and the increase in error after the minimum began to steepen.

Having a large single-photon detuning, $\Delta$, is not always easy to achieve depending on the experimental setup as it requires large laser power. With this in mind, we studied how the large single-photon detuning would play a role in the error FOM. We looked at large detunings ranging from $20$ $\mathrm{GHz}$ to $1$ $\mathrm{GHz}$. We saw that as the detuning increased, the number of pulses to get a minimum FOM increased. For smaller detunings, the number of pulses for minimum $\mathrm{FOM}$ decreased as well. There was a steeper rise in the DC offset error once this minimum occurred for smaller detunings due to the higher probability of spontaneous decay. As expected in other literature \cite{CSP09}, we quantitatively analyzed that it is preferred to have a larger single-photon detuning to limit the amount of spontaneous decay as shown here, but the increase in intensity effect on the contrast of the atom interferometer will need to be considered.

The $\Delta$ detuning gives a limit on the amount of spontaneous decay, but in an experiment it is not always easy, or desired, to have the two-photon detuning $\delta=0$. When considering systems with $\Delta$ and $\delta$ values similar to the results from Butts \textit{et al} \cite{BKK13} and McGuirk \textit{et al} \cite{MSK00}, we see that the $\delta$ detuning starts limiting the atom interferometer performance appreciably. For $\delta$ values used by Butts \textit{et al} and considering plausible Doppler and recoil shifts, the increase in $\Delta$ gave less of an increase in overall performance compared to the $\delta =0$ case. This is due to $\delta$ affecting the efficiency of the $\pi$-pulse more directly. The less efficient pulses will have the DC offset error affected by two-factors instead of just the quantum information loss. We also came in good agreement with the amount of quantum information loss per pulse for the $6\hbar k_\mathrm{eff}$ and $10\hbar k_\mathrm{eff}$ cases to that seen by Butts \textit{et al} (e.g. $0.26\%$ vs $0.25\%$) \cite{BKK13}. This unfortunately was not the case when comparing to McGuirk \textit{et al}. The discrepancy in our per pulse quantum information loss to their estimated value comes from not knowing the correct $\delta$ two-photon detuning in their experiment as well as the difference in the atomic system ($^{85}$Rb compared to $^{133}$Cs).

Our results show that the spontaneous decay of LMT multiple-Raman pulses will have a limit where the error will be minimum based on different experimental parameters. Depending on the large single-photon detuning and uncertainty in the measurement of the excited state, a LMT of $34\hbar k_\mathrm{eff}$ or higher can be achieved with minimum deviation in measurement before spontaneous decay starts to be an important factor. At the optimal LMT of $34\hbar k_\mathrm{eff}$, we observed the overall error on the order of $10^{-5}$ $m/s^{2}$. This could be the reason why experiments utilizing multi-Raman pulse LMT combined it with other error-compensating techniques \cite{BKK13,KBK15}. For instance, the experiments using composite Raman pulses on $^{133}$Cs which saw a minimal loss of contrast due to spontaneous decay \cite{BKK13}. Kotru \textit{et al} even resolved an LMT of $30\hbar k_\mathrm{eff}$ when using a combination of multiple-Raman pulses and frequency-swept adiabatic rapid passage \cite{KBK15}. With this in mind, these experimental insights give plausibility to these LMT amounts when using multiple-Raman pulse techniques. Finally, we note that we have considered mainly the effect of spontaneous decay in this work, but other noise factors (i.e. laser phase noise, shot noise, etc. \cite{CCD05,GMK08,CSP09,RPV14}) will also affect the performance of LMT atom interferometers using multiple-Raman pulses, which are out of the scope of the work presented here. We also consider an ideal situation where the $D_{2}$ transition of the $\left\vert F = 2, m_{f}=0 \right\rangle$ ground state to the $\left\vert F^{\prime} = 2 \right\rangle$ intermediate state is suppressed. Future work could incorporate this transition into the dynamics as well to see how much more information loss there will be.

\begin{acknowledgments}
Sandia National Laboratories is a multi-mission laboratory managed and operated by National Technology and Engineering Solutions of Sandia, LLC, a wholly-owned subsidiary of Honeywell International, Inc., for the DOE's National Nuclear Security Administration under contract DE-NA0003525. This paper describes objective technical results and analysis. Any subjective views or opinions that might be expressed in the paper do not necessarily represent the views of the U.S. Department of Energy or the United States Government.
\end{acknowledgments}

\begin{appendix}

\section{Full Open Quantum System Dynamics} \label{AppendixA}
\subsection{State population dynamics} \label{statepopdynamics}
Here we will discuss the solution to the adiabatically eliminated state $c_i(t)$ as discussed in the main text \ref{sec_dynamics}. Let us show again the dynamics of the three-level system when considering our effective Hamiltonian and the Schrodinger equation,%
\begin{widetext}
\begin{subequations}
\begin{eqnarray}
i\hbar\overset{\cdot}{c}_{g}&=&\hbar\Omega_{1}e^{i(-\mathbf{k}_{1}\cdot
x_{g}+\omega_{1}t)}c_{i}, \\
i\hbar\overset{\cdot}{c}_{e}&=&\hbar\Omega_{2}e^{i(-\mathbf{k}_{2}\cdot
x_{e}+\omega_{2}t)}c_{i}+\hbar\omega_{ge}c_{e},\\
i\hbar\overset{\cdot}{c}_{i}&=&\hbar\Omega_{1}^{\ast}e^{i(\mathbf{k}_{1}\cdot\mathbf{x}%
_{g}-\omega_{1}t)}{c}_{g} + \hbar\Omega_{2}^{\ast}e^{i(\mathbf{k}_{2}\cdot\mathbf{x}_{e}-\omega_{2}t)}{c}_{e}+\hbar\omega_{gi}c_{i}-i\hbar
\frac{\gamma_{l}}{2}c_{i}.
\end{eqnarray}
\end{subequations}
\end{widetext}
We next move into the rotating frame and use the substitutions $c_{g}=\widetilde{c}_{g}$, $c_{e}=\widetilde{c}_{e}e^{-i\omega_{ge}t}$, and $c_{i}=\widetilde{c}_{i}e^{-i\omega_{gi}t}$ to help simplify solving this set of coupled differential equations, and divide by $\hbar$ to get a slightly different differential equation for
$\overset{\cdot}{\widetilde{c}_{i}}$. We will also use that the dynamics for $\widetilde{c}_{i}$ will be very fast when compared to $\widetilde{c}_{g}$ and $\widetilde{c}_{e}$ allowing us to use treat them as constant and adiabatically eliminate $\widetilde{c}_{i}$ \cite{BPM07,FVS20}. We now go about solving the differential equation of $\widetilde{c}_{i}$,%
\begin{widetext}
\begin{subequations}

\begin{align}
i\frac{d\left(\widetilde{c}_{i}e^{-i\omega_{gi}t}\right)}{dt}%
&=\Omega_{1}^{*}e^{-i\left(k_{1}\cdot x_{g}-\omega_{1}t\right)}\widetilde{c}_{g}+\Omega_{2}^{*}e^{-i\left(k_{2}\cdot x_{e}-\left(\omega_{2}+\omega{eg}\right)t\right)}\widetilde{c}_{e}+\omega_{gi}\widetilde{c}_{i}e^{-i\omega_{gi}t}-i\frac{\gamma_{l}}{2}\widetilde{c}_{i}e^{-i\omega_{gi}t}, \\  
i\frac{d\left( \widetilde{c}_{i}\right)  }{dt}e^{-i\omega_{gi}t}+\omega_{gi}\widetilde{c}_{i}e^{-i\omega_{gi}t}&=\Omega_{1}^{*}e^{-i\left(k_{1}\cdot x_{g}-\omega_{1}t\right)}\widetilde{c}_{g}+\Omega_{2}^{*}e^{-i\left(k_{2}\cdot x_{e}-\left(\omega_{2}+\omega{eg}\right)t\right)}\widetilde{c}_{e}+\omega_{gi}\widetilde{c}_{i}e^{-i\omega_{gi}t}-i\frac{\gamma_{l}}{2}\widetilde{c}_{i}e^{-i\omega_{gi}t}, \\
\frac{d\left(  \widetilde{c}_{i}\right)  }{dt}+\frac{\gamma_{l}}{2}\widetilde{c}%
_{i}&=-i\Omega_{1}^{\ast}e^{-i(\mathbf{k}_{1}\cdot\mathbf{x}_{g}+\Delta
t)}\widetilde{c}_{g}-i\Omega_{2}^{\ast}e^{-i(\mathbf{k}_{2}\cdot\mathbf{x}%
_{e}+(\Delta+\delta)t)}\widetilde{c}_{e}. \label{Newcidynamics}%
\end{align}
\end{subequations}
\end{widetext}
Here we have used the definitions $\Delta=\omega_{gi}-$ $\omega_{1}$ and
$\delta=\omega_{1}-\omega_{2}-\omega_{ge}$. This has our dynamics be dependent
on the two detuning parameters and the spontaneous decay as expected. Since we can apply adiabatic elimination, we treat $\widetilde{c}_{g}$ and $\widetilde{c}_{e}$ as constants when compared to the extremely fast dynamics of $\widetilde{c}_{i}$.

To solve the differential equation (Eqn. (\ref{Newcidynamics})) for the
dynamics of $\widetilde{c}_{i}$ let us consider that $a(t)=\gamma_{l}/2$ and the
right-hand side is $b(t)$. This gives a differential equation of the form%
\begin{equation}
\frac{d\left(\widetilde{c}_{i}\right)}{dt}+a(t)\widetilde{c}_{i}=b(t).
\end{equation}
This differential equation has a general solution given by%
\begin{equation}
\widetilde{c}_{i}(t)=Ce^{-Y(t)}+e^{-Y(t)}\int e^{Y(t^{^{\prime}})}b(t^{^{\prime}})dt^{^{\prime}}. \label{DiffEqnGenSol}%
\end{equation}
Here $Y(t)=\int\frac{\gamma_{l}}{2}dt^{^{\prime}}$ and we can plug back in $a(t)=\gamma_{3}
/2$ to get $Y(t)=\frac{\gamma_{l}}{2}t$. We then plug $Y(t)$ into Eqn.
(\ref{DiffEqnGenSol}) to get
\begin{widetext}
\begin{equation}
\widetilde{c}_{i}=Ce^{-\frac{\gamma_{l}}{2}t}+e^{-\frac{\gamma_{l}}{2}t}\int
e^{\frac{\gamma_{l}}{2}t^{\prime}}\left(-i\Omega_{1}^{\ast}e^{-i(\mathbf{k}_{1}\cdot
\mathbf{x}_{g}+\Delta t)}\widetilde{c}_{g}-i\Omega_{2}^{\ast}e^{-i(\mathbf{k}%
_{2}\cdot\mathbf{x}_{e}+(\Delta+\delta)t)}\widetilde{c}_{e}\right)dt^{^{\prime}},
\end{equation}
\end{widetext}
It can be seen that if $\gamma_{l}=0$, we return to the usual closed quantum system
without the non-Hermitian spontaneous decay. Performing the integration, we
get
\begin{widetext}
\begin{align}
\widetilde{c}_{i}(t)  &  =Ce^{-\frac{\gamma_{l}}{2}t}+\left[  \frac{-i\Omega_{1}^{\ast}}{\frac{\gamma_{l}}{2}-i\Delta}e^{i(\mathbf{k}_{1}\cdot\mathbf{x}_{g}+\Delta t)}\widetilde{c}_{g}+\frac{-i\Omega_{2}^{\ast}%
}{\frac{\gamma_{l}}{2}-i(\Delta+\delta)}e^{i(\mathbf{k}_{2}\cdot\mathbf{x}%
_{e}+(\Delta+\delta)t)}\widetilde{c}_{e}\right] \label{Newcipop}
\end{align}
\end{widetext}
Next, we need to determine the value for $C$. To do this we consider the initial condition that our system will be completely in the ground state (i.e.
$\widetilde{c}_{i}(t=0)=0$) which gives us
\begin{equation}
C=\left[  \frac{i\Omega_{1}^{\ast}}{i\Delta-\frac{\gamma_{l}}{2}}e^{i(\mathbf{k}%
_{1}\cdot\mathbf{x}_{g})}\widetilde{c}_{g}+\frac{i\Omega_{2}^{\ast}}%
{i(\Delta+\delta)-\frac{\gamma_{l}}{2}}e^{i(\mathbf{k}_{2}\cdot\mathbf{x}_{e}%
)}\widetilde{c}_{e}\right]
\end{equation}
Plugging this back into Eqn. (\ref{Newcipop}), the final solution for $\widetilde{c}_{i}$ will be,
\begin{widetext}
\begin{align}
\widetilde{c}_{i}  &  =\left[  \frac{i\Omega_{1}^{\ast}}{i\Delta-\frac{\gamma_{l}}{2}}e^{i(\mathbf{k}_{1}\cdot\mathbf{x}_{g})}\widetilde{c}_{g}+\frac
{i\Omega_{2}^{\ast}}{i(\Delta+\delta)-\frac{\gamma_{l}}{2}}e^{i(\mathbf{k}%
_{2}\cdot\mathbf{x}_{e})}\widetilde{c}_{e}\right]  e^{-\frac{\gamma_{l}}{2}%
t} -\left[  \frac{i\Omega_{1}^{\ast}}{i\Delta-\frac{\gamma_{l}}{2}}e^{i(\mathbf{k}%
_{1}\cdot\mathbf{x}_{g}+\Delta t)}\widetilde{c}_{g}+\frac{i\Omega_{2}^{\ast}%
}{i(\Delta+\delta)-\frac{\gamma_{l}}{2}}e^{i(\mathbf{k}_{2}\cdot\mathbf{x}_{e}+(\Delta+\delta)t)}\widetilde{c}_{e}\right] \label{cipopFinal}
\end{align}
\end{widetext}
With our solution for the intermediate high energy state, we can plug Eqn.
(\ref{cipopFinal}) back into our dynamics for $\widetilde{c}_{g}$ and
$\widetilde{c}_{e}$. Doing the algebra will leave us with,
\begin{widetext}
\begin{subequations}
\begin{eqnarray}
\overset{\cdot}{\widetilde{c}}_{g}  &  = &\left[  \frac{\left\vert \Omega_{1}\right\vert^{2}}{i\Delta-\frac{\gamma_{l}}{2}}e^{i\Delta t}\widetilde{c}_{g}+\frac{\Omega_{1}\Omega_{2}^{\ast}}{i(\Delta+\delta)-\frac{\gamma_{l}}{2}}e^{i(\Delta_{kx}-\Delta t)}\widetilde{c}_{e}\right]e^{-\frac{\gamma_{l}}{2}t}
-\left[  \frac{\left\vert \Omega_{1}\right\vert ^{2}}{i\Delta-\frac{\gamma_{l}}{2}}\widetilde{c}_{g}+\frac{\Omega_{1}\Omega_{2}^{\ast}}{i(\Delta+\delta)-\frac{\gamma_{l}}{2}}e^{i(\Delta_{kx}-\delta t)}\widetilde{c}_{e}\right]
,\label{NewcgDynamicsApp}\\
\overset{\cdot}{\widetilde{c}}_{e}  &  = &\left[  \frac{\left\vert \Omega_{2}\right\vert
^{2}}{i(\Delta+\delta)-\frac{\gamma_{l}}{2}}e^{i(\Delta+\delta)t}\widetilde{c}_{e}+\frac{\Omega_{2}\Omega_{1}^{\ast}}{i\Delta-\frac{\gamma_{l}}{2}}e^{i(-\Delta_{kx}-(\Delta+\delta)t)}\widetilde{c}_{g}\right] e^{-\frac{\gamma_{l}}{2}t} \nonumber\\
&&\hspace{6.0cm}-\left[  \frac{\left\vert \Omega_{2}\right\vert ^{2}}{i(\Delta+\delta)-\frac{\gamma_{l}}{2}}\widetilde{c}_{e}+\frac{\Omega_{2}\Omega_{1}^{\ast}%
}{i\Delta-\frac{\gamma_{l}}{2}}e^{i(-\Delta_{kx}-\delta t)}%
\widetilde{c}_{g}\right]  .\label{NewceDynamicsApp}
\end{eqnarray}
\end{subequations}
\end{widetext}

\subsection{Density matrix dynamics} \label{densmatrixdyn}
\subsubsection{Closed system dynamics} \label{closeddensmat}
So far we have considered the dynamics for the amplitudes for the atomic states. Another way to describe the dynamics is through the density matrix. Let us start by considering again the following equation, 
\begin{equation}
\overset{\cdot}{\rho}=-\frac{i}{\hbar}\left[\widehat{H},\widehat{\rho}\right]  +\sum_{k}\left(L_{k}\rho L_{k}^{\dagger}-\frac{1}{2} L_{k}^{\dagger}L_{k}\rho - \frac{1}{2} \rho L_{k}^{\dagger}L_{k}\right) .
\end{equation}
Where we are using that $\widehat{H}=\widehat{H}_{0}+\widehat{H}_{\mathrm{int}}$ and $L_{k}$ are our Lindblad jump operators same as before. We start by considering a closed system where we only will have $L_{g}$ and $L_{e}$. We now can get our dynamics for each element of the density matrix,
\begin{widetext}
\begin{subequations}
\begin{align}
\overset{\cdot}{\rho}_{gg} &= \Omega_{1}e^{i(k_{1}x_{g}-\omega_{1}t)}\rho_{gi}+\Omega_{1}^{\ast}e^{-i(k_{1}x_{g}-\omega_{1}t)}\rho_{ig}+\gamma_{g}\rho_{ii} , \\
\overset{\cdot}{\rho}_{gi} &= \overset{\cdot}{\rho}^{\ast}_{ig} =  i\omega_{ig}\rho_{gi} + \Omega_{1}^{\ast}e^{-i(k_{1}x_{g}-\omega_{1}t)}(\rho_{ii}-\rho_{gg}) - \Omega_{2}^{\ast}e^{-i(k_{2}x_{e}-\omega_{2}t)}\rho_{ge} - \frac{\gamma_{g}+\gamma_{e}}{2}\rho_{gi} , \\
\overset{\cdot}{\rho}_{ee} &= \Omega_{2}^{\ast}e^{-i(k_{2}x_{e}-\omega_{2}t)}\rho_{ie} +\Omega_{2}e^{i(k_{2}x_{e}-\omega_{2}t)}\rho_{ei} + \gamma_{e}\rho_{ii} ,\\
\overset{\cdot}{\rho}_{ei} &= \overset{\cdot}{\rho}^{\ast}_{ie} = i\omega_{ie}\rho_{ei} - \Omega_{1}^{\ast}e^{-i(k_{1}x_{g}-\omega_{1}t)}\rho_{eg} - \Omega_{2}^{\ast}e^{-i(k_{2}x_{e}-\omega_{2}t)}(\rho_{ii}-\rho_{ee}) - \frac{\gamma_{g}+\gamma_{e}}{2}\rho_{ei} ,\\
\overset{\cdot}{\rho}_{eg} &= \overset{\cdot}{\rho}^{\ast}_{ge} = i\omega_{eg}\rho_{eg} + \Omega_{1}^{\ast}e^{-i(k_{1}x_{g}-\omega_{1}t)}\rho^{\ast}_{ie} + \Omega_{2}e^{i(k_{2}x_{e}-\omega_{2}t)}\rho_{ig},\\
\overset{\cdot}{\rho}_{ii} &= -\overset{\cdot}{\rho}_{gg} - \overset{\cdot}{\rho}_{ee}.
\end{align}
\end{subequations}
\end{widetext}
We can simplify these equations by moving into the rotating frame just like we did for the amplitude dynamics,
\begin{subequations}
\begin{align}
\rho_{kk} &= \widetilde{\rho}_{kk} , \\
\rho_{ge} &= \rho_{eg}^{\ast} = \widetilde{\rho}_{ge}e^{i(\omega_{1}-\omega_{2})t} , \\
\rho_{gi} &= \rho_{ig}^{\ast} = \widetilde{\rho}_{ig}e^{i\omega_{1}t} , \\
\rho_{ei} &= \rho_{ie}^{\ast} = \widetilde{\rho}_{ei}e^{i\omega_{2}t} .
\label{Eq:Transformation}
\end{align}
\end{subequations}
Next, we will consider a spontaneous emission time scale as $\tau = \Gamma t$ where $\Gamma$ is the spontaneous emission rate and is only considering decay from state $\left\vert i\right\rangle$ to states $\left\vert g\right\rangle$ and $\left\vert e\right\rangle$. Here we will make an assumption that the decay into states $\left\vert g\right\rangle$ and $\left\vert e\right\rangle$ are equal. We can then make the following dimensionless transformations starting with,
\begin{align}
\frac{d}{dt} &= \Gamma \frac{d}{d\tau}, \\
\widetilde{\Omega} &= \frac{\Omega}{\Gamma}.
\end{align}
This will now have our dimensionless dynamics of the closed system to be,
\begin{widetext}
\begin{subequations}
\begin{align}
\overset{\cdot}{\widetilde{\rho}}_{gg} &= \widetilde{\Omega}_{1}\widetilde{\rho}_{gi}+\widetilde{\Omega}_{1}^{\ast}\widetilde{\rho}_{ig}+\frac{1}{2}\widetilde{\rho}_{ii} , \\
\overset{\cdot}{\widetilde{\rho}}_{gi} &= \overset{\cdot}{\widetilde{\rho}^{\ast}_{ig}} =  \frac{i}{\Gamma}( \omega_{ig} - (\omega_{1}-\vec{k}_{1}\cdot \vec{v}_{g}))\widetilde{\rho}_{gi} + \widetilde{\Omega}_{1}^{\ast}(\widetilde{\rho}_{ii}-\widetilde{\rho}_{gg}) - \widetilde{\Omega}_{2}^{\ast}\widetilde{\rho}_{ge} - \frac{1}{2}\widetilde{\rho}_{gi} , \\
\overset{\cdot}{\widetilde{\rho}}_{ee} &= \widetilde{\Omega}_{2}^{\ast}\widetilde{\rho}_{ie} +\widetilde{\Omega}_{2}\widetilde{\rho}_{ei} + \frac{1}{2}\widetilde{\rho}_{ii} ,\\
\overset{\cdot}{\widetilde{\rho}}_{ei} &= \overset{\cdot}{\widetilde{\rho}^{\ast}}_{ie} = \frac{i}{\Gamma}(\omega_{ie}-(\omega_{2}-\vec{k}_{2}\cdot \vec{v}_{e}))\widetilde{\rho}_{ei} - \widetilde{\Omega}_{1}^{\ast}\rho_{eg} - \widetilde{\Omega}_{2}^{\ast}(\widetilde{\rho}_{ii}-\widetilde{\rho}_{ee}) - \frac{1}{2}\widetilde{\rho}_{ei} ,\\
\overset{\cdot}{\widetilde{\rho}}_{eg} &= \overset{\cdot}{\widetilde{\rho}^{\ast}_{ge}} = \frac{i}{\Gamma}(\omega_{eg}-(\omega_{1}-\vec{k}_{1}\cdot \vec{v}_{g})-(\omega_{2}-\vec{k}_{2}\cdot \vec{v}_{e}))\widetilde{\rho}_{eg} + \widetilde{\Omega}_{1}^{\ast}\widetilde{\rho^{\ast}}_{ie} + \widetilde{\Omega}_{2}\widetilde{\rho}_{ig},\\
\overset{\cdot}{\widetilde{\rho}}_{ii} &= -\overset{\cdot}{\widetilde{\rho}}_{gg} - \overset{\cdot}{\widetilde{\rho}}_{ee}.
\end{align}
\end{subequations}
\end{widetext}
The Doppler-shifted laser frequencies observed by the atoms are represented by the $\omega - \vec{k} \cdot \vec{v}$ terms. Thus far we have not taken into account the Doppler shift as we have been interested in looking at the spontaneous emission. With that in mind, we will ignore its contribution even for the closed system dynamics and make the following substitutions similar to what was done previously,
\begin{subequations}
\begin{align}
\Delta &= \omega_{ig}-\omega_{1}, \\
\delta_{R} &= \omega_{eg}-(\omega_{1}-\omega_{2}).
\end{align}
\end{subequations}
This will give us reduced dynamics of the following,
\begin{widetext}
\begin{subequations}
\begin{align}
\overset{\cdot}{\widetilde{\rho}}_{gg} &= \widetilde{\Omega}_{1}\widetilde{\rho}_{gi}+\widetilde{\Omega}_{1}^{\ast}\widetilde{\rho}_{ig}+\frac{1}{2}\widetilde{\rho}_{ii} , \\
\overset{\cdot}{\widetilde{\rho}}_{gi} &= \overset{\cdot}{\widetilde{\rho}^{\ast}_{ig}} =  i\widetilde{\Delta}\widetilde{\rho}_{gi} + \widetilde{\Omega}_{1}^{\ast}(\widetilde{\rho}_{ii}-\widetilde{\rho}_{gg}) - \widetilde{\Omega}_{2}^{\ast}\widetilde{\rho}_{ge} - \frac{1}{2}\widetilde{\rho}_{gi} , \\
\overset{\cdot}{\widetilde{\rho}}_{ee} &= \widetilde{\Omega}_{2}^{\ast}\widetilde{\rho}_{ie} +\widetilde{\Omega}_{2}\widetilde{\rho}_{ei} + \frac{1}{2}\widetilde{\rho}_{ii} ,\\
\overset{\cdot}{\widetilde{\rho}}_{ei} &= \overset{\cdot}{\widetilde{\rho}^{\ast}_{ie}} = i(\widetilde{\Delta}-\widetilde{\delta_{R}})\widetilde{\rho}_{ei} - \widetilde{\Omega}_{1}^{\ast}\rho_{eg} - \widetilde{\Omega}_{2}^{\ast}(\widetilde{\rho}_{ii}-\widetilde{\rho}_{ee}) - \frac{1}{2}\widetilde{\rho}_{ei} ,\\
\overset{\cdot}{\widetilde{\rho}}_{eg} &= \overset{\cdot}{\widetilde{\rho}^{\ast}_{ge}} = i\widetilde{\delta_{R}}\widetilde{\rho}_{eg} + \widetilde{\Omega}_{1}^{\ast}\widetilde{\rho^{\ast}}_{ie} + \widetilde{\Omega}_{2}\widetilde{\rho}_{ig},\\
\overset{\cdot}{\widetilde{\rho}}_{ii} &= -\overset{\cdot}{\widetilde{\rho}}_{gg} - \overset{\cdot}{\widetilde{\rho}}_{ee}.
\end{align}
\end{subequations}
\end{widetext}
Here $\widetilde{\Delta}=\Delta / \Gamma$ and $\widetilde{\delta_{R}}=\delta_{R} / \Gamma$.

The dynamics above is a closed system where the high energy excited state $\left\vert i\right\rangle$ emits population into either $\left\vert g\right\rangle$ or $\left\vert e\right\rangle$ giving no loss of population in the system, $\rho_{ii}+\rho_{ee}+\rho_{gg}=1$. For the open quantum system, we will consider that the state $\left\vert i\right\rangle$ will decay into something other than $\left\vert g\right\rangle$ and $\left\vert e\right\rangle$. This spontaneous emission is added in through the following,
\begin{subequations}
\begin{align}
\overset{\cdot}{\widetilde{\rho}}_{ii} &= -\overset{\cdot}{\widetilde{\rho}}_{gg} - \overset{\cdot}{\widetilde{\rho}}_{ee} - \widetilde{\gamma_{l}}\widetilde{\rho}_{ii},
\end{align}
\end{subequations}
where $\widetilde{\gamma_{l}}=\gamma_{l}/\Gamma$.

\subsubsection{Open system dynamics}
Another way to look at the dynamics is to start by considering it an open quantum system from the start. Let us return to the Hamiltonian for the atom interferometer three-level system and apply the Lindblad operators for each decay channel to determine $\overset{\cdot}{\rho}$ for the system. Using equation (\ref{Eq:Lindblad}) and with the help of equation (\ref{Eq:Transformation}) we find a set of differential equations for the density matrix components as follows \cite{SAD14},
\begin{widetext}
\begin{subequations}
\begin{align}
\overset{\cdot}{\widetilde{\rho}}_{gg} &= i\Omega_{1}e^{ik_{1}x_{g}}\widetilde{\rho}_{gi}-i\Omega_{1}^{\ast}e^{-ik_{1}x_{g}}\widetilde{\rho}_{ig}+\gamma_{g}\widetilde{\rho}_{ii} , \\
\overset{\cdot}{\widetilde{\rho}}_{ge} &= \overset{\cdot}{\widetilde{\rho}^{\ast}_{eg}} = -i\delta\widetilde{\rho}_{ge} - i\Omega_{1}^{\ast}e^{-ik_{1}x_{g}}\widetilde{\rho}_{ie} + i\Omega_{2}e^{ik_{2}x_{e}}\widetilde{\rho}_{gi} , \\
\overset{\cdot}{\widetilde{\rho}}_{gi} &= \overset{\cdot}{\widetilde{\rho}^{\ast}_{ig}} =  i\Delta\widetilde{\rho}_{gi} - i[\Omega_{1}^{\ast}e^{-ik_{1}x_{g}}(\widetilde{\rho}_{ii}+\widetilde{\rho}_{ee}-1) - \Omega_{2}^{\ast}e^{-ik_{2}x_{e}}\widetilde{\rho}_{ge}] - \frac{\Gamma}{2}\widetilde{\rho}_{gi} , \\
\overset{\cdot}{\widetilde{\rho}}_{ee} &= -i\Omega_{2}^{\ast}e^{-ik_{2}x_{e}}\widetilde{\rho}_{ie} +i\Omega_{2}e^{ik_{2}x_{e}}\widetilde{\rho}_{ei} + \gamma_{e}\widetilde{\rho}_{ii} ,\\
\overset{\cdot}{\widetilde{\rho}}_{ei} &= \overset{\cdot}{\widetilde{\rho}_{ie}^{\ast}} = i(\Delta+\delta)\widetilde{\rho}_{ei} - i[\Omega_{1}^{\ast}e^{-ik_{1}x_{g}}\widetilde{\rho}_{eg} - \Omega_{2}^{\ast}e^{-ik_{2}x_{e}}(\widetilde{\rho}_{ii}-\widetilde{\rho}_{ee})] - \frac{\Gamma}{2}\widetilde{\rho}_{ei} ,\\
\overset{\cdot}{\widetilde{\rho}}_{ii} &= -\overset{\cdot}{\widetilde{\rho}}_{gg} - \overset{\cdot}{\widetilde{\rho}}_{ee} - \gamma_{l}\widetilde{\rho}_{ii}.
\end{align}
\end{subequations}
\end{widetext}
Here $\Gamma= \gamma_{g}+\gamma_{e}+\gamma_{l}$. We now use the fact that we are far detuned from the intermediate excited state (i.e. $\Delta \gg \delta,\Omega_{1},\Omega_{2}, \gamma_{g,e,l}$) to adiabatically eliminate the intermediate excited state. This adiabatic elimination brings a fast decay of the quickly oscillating term for the $\overset{\cdot}{\widetilde{\rho}}_{gi}$ and $\overset{\cdot}{\widetilde{\rho}}_{ei}$ dynamics. We also will have the population of the intermediate excited state, $\widetilde{\rho}_{ii}$, go to zero from the depumping and fast spontaneous decay. With this, we can solve for their steady-state solutions to get

\begin{subequations}
\begin{align}
\widetilde{\rho}_{gi}^{SS} &= i\frac{\Omega_{1}^{\ast}\widetilde{\rho}_{gg}e^{-k_{1}x_{g}} - \Omega_{2}^{\ast}\widetilde{\rho}_{ge}e^{-k_{2}x_{e}}}{(\frac{\Gamma}{2}-i\Delta)}, \label{rho13SS} \\ 
\widetilde{\rho}_{ei}^{SS} &= -i\frac{\Omega_{1}^{\ast}\widetilde{\rho}_{eg}e^{-k_{1}x_{g}} + \Omega_{2}^{\ast}\widetilde{\rho}_{ee}e^{-k_{2}x_{e}}}{(\frac{\Gamma}{2}-i(\Delta+\delta))} . \label{rho23SS}
\end{align}
\end{subequations}

With the steady-state solutions for the slowly varying terms, we can plug Eqns. (\ref{rho13SS}) and (\ref{rho23SS}) back into $\overset{\cdot}{\widetilde{\rho}}_{gg}$, $\overset{\cdot}{\widetilde{\rho}}_{ge}$, and $\overset{\cdot}{\widetilde{\rho}}_{ee}$ and use the fact to that $\Delta \gg \delta$ to get a set of coupled differential equations that depend only on the $|g \rangle$ and $|e \rangle$ states
\begin{widetext}
\begin{subequations}
\begin{align}
\overset{\cdot}{\widetilde{\rho}}_{gg} &= |\Omega_{1}|^{2}\widetilde{\rho}_{gg}\left[\frac{1}{\frac{\Gamma}{2}-i\Delta}+\frac{1}{\frac{\Gamma}{2}+i\Delta}\right] + \frac{\Omega_{1}\Omega_{2}^{\ast}}{\frac{\Gamma}{2}-i\Delta}e^{i\Delta_{kx}}\widetilde{\rho}_{ge} + \frac{\Omega_{2}\Omega_{1}^{\ast}}{\frac{\Gamma}{2}+i\Delta}e^{-i\Delta_{kx}}\widetilde{\rho}_{eg} , \label{rho11dynamics} \\
\overset{\cdot}{\widetilde{\rho}}_{ge} &= -\frac{\Omega_{1}^{\ast}\Omega_{2}}{\frac{\Gamma}{2}+i\Delta}e^{-i\Delta_{kx}}\widetilde{\rho}_{ee} - \frac{\Omega_{1}^{\ast}\Omega_{2}}{\frac{\Gamma}{2}-i\Delta}e^{i\Delta^{'}_{kx}}\widetilde{\rho}_{gg} - \left[\frac{|\Omega_{1}|^{2}}{\frac{\Gamma}{2}+i\Delta}+\frac{|\Omega_{2}|^{2}}{\frac{\Gamma}{2}-i\Delta} \right]\widetilde{\rho}_{ge}, \label{rho12dynamics} \\
\overset{\cdot}{\widetilde{\rho}}_{ee} &= -|\Omega_{2}|^{2}\widetilde{\rho}_{ee}\left[\frac{1}{i\Delta+\frac{\Gamma}{2}}+\frac{1}{\frac{\Gamma}{2}-i\Delta}\right]-\frac{\Omega_{2}^{\ast}\Omega_{1}}{\frac{\Gamma}{2}+i\Delta}e^{-i\Delta^{'}_{kx}}\widetilde{\rho}_{ge}-\frac{\Omega_{1}^{\ast}\Omega_{2}}{\frac{\Gamma}{2}-i\Delta}e^{i\Delta^{'}_{kx}}\widetilde{\rho}_{eg}. \label{rho22dynamics}
\end{align}
\end{subequations}
\end{widetext}
We next want to see if our initial model equations will give us the same outcome when we consider that $\rho_{kj}=\langle k|\rho|j \rangle = \rho_{kj}=\langle k|\psi\rangle \langle\psi|j \rangle = c_{k}^{\ast}c_{j}$. The differential form will have $\overset{\cdot}{\rho}_{kj} = d({c}^{\ast}_{k}{c}_{j})/dt$. Using Eqns. (\ref{NewcgDynamicsApp}) and (\ref{NewceDynamicsApp}), replacing $\gamma_{l}$ with $\Gamma$, applying the chain rule, doing some algebra will give us
\begin{widetext}
\begin{subequations}
\begin{align}
\frac{d(c_{g}^{\ast}c_{g})}{dt} &= -|\Omega_{1}|^{2}\widetilde{c^{\ast}_{g}}\widetilde{c}_{g}\left[\frac{f(\Delta,\Gamma)}{\frac{\Gamma}{2}-i\Delta}+\frac{f^{\ast}(\Delta,\Gamma)}{\frac{\Gamma}{2}+i\Delta}\right]- \frac{\Omega_{1}\Omega_{2}^{\ast}}{\frac{\Gamma}{2}-i\Delta}\widetilde{c^{\ast}_{g}}\widetilde{c}_{e}e^{i\Delta_{kx}}f^{\ast}(\Delta,\Gamma) - \frac{\Omega_{2}\Omega_{1}^{\ast}}{\frac{\Gamma}{2}+i\Delta}\widetilde{c^{\ast}_{e}}\widetilde{c}_{g}e^{-i\Delta_{kx}}f(\Delta,\Gamma) , \\
\frac{d(c_{g}^{\ast}c_{e})}{dt} &= \frac{\Omega_{1}^{\ast}\Omega_{2}}{\frac{\Gamma}{2}+i\Delta}\widetilde{c^{\ast}_{e}}\widetilde{c}_{e}e^{-i\Delta_{kx}}f(\Delta,\Gamma)+\frac{\Omega_{1}^{\ast}\Omega_{2}}{\frac{\Gamma}{2}+i\Delta}\widetilde{c^{\ast}_{g}}\widetilde{c}_{g}e^{i\Delta^{'}_{kx}}f^{\ast}(\Delta,\Gamma) + \left[ \frac{|\Omega_{1}|^{2}}{\frac{\Gamma}{2}+i\Delta}f(\Delta,\Gamma)+\frac{|\Omega_{2}|^{2}}{\frac{\Gamma}{2}-i\Delta}f^{\ast}(\Delta,\Gamma) \right]\widetilde{c^{\ast}_{g}}\widetilde{c}_{e}\\
\frac{d(c_{e}^{\ast}c_{e})}{dt} &= |\Omega_{2}|^{2}\widetilde{c^{\ast}_{e}}\widetilde{c}_{e}\left[\frac{f^{\ast}(\Delta,\Gamma)}{i\Delta+\frac{\Gamma}{2}}+\frac{f(i\Delta,\frac{\Gamma}{2})}{\frac{\Gamma}{2}-i\Delta}\right] +\frac{\Omega_{2}^{\ast}\Omega_{1}}{\frac{\Gamma}{2}+i\Delta}\widetilde{c^{\ast}_{g}}\widetilde{c}_{e}e^{-i\Delta^{'}_{kx}}f(\Delta,\Gamma)+\frac{\Omega_{1}^{\ast}\Omega_{2}}{\frac{\Gamma}{2}-i\Delta}\widetilde{c^{\ast}_{e}}\widetilde{c}_{g}e^{i\Delta^{'}_{kx}}f^{\ast}(\Delta,\Gamma). 
\end{align}
\end{subequations}
\end{widetext}
Here the function $f(\Delta,\Gamma)=e^{i\Delta t - (\Gamma/2)t}-1$ and $f^{\ast}(\Delta,\Gamma)$ is the complex conjugate. It can be seen that the two methods match up almost exactly. The only differences come from an additional quickly oscillating and decaying term and a sign difference. We again can consider that the oscillations will be extremely fast and that the decay will be fast as well (i.e. adiabatic elimination). This eliminates the quickly oscillating and decaying from the $f(i\Delta,\frac{\Gamma}{2})$ term and gives us a sign change
\begin{widetext}
\begin{subequations}
\begin{align}
\frac{d(c_{g}^{\ast}c_{g})}{dt} &= |\Omega_{1}|^{2}\widetilde{c^{\ast}_{g}}\widetilde{c}_{g}\left[\frac{1}{\frac{\Gamma}{2}-i\Delta}+\frac{1}{\frac{\Gamma}{2}+i\Delta}\right] + \frac{\Omega_{1}\Omega_{2}^{\ast}}{\frac{\Gamma}{2}-i\Delta}e^{i\Delta_{kx}}\widetilde{c^{\ast}_{g}}\widetilde{c}_{e} + \frac{\Omega_{2}\Omega_{1}^{\ast}}{\frac{\Gamma}{2}+i\Delta}e^{-i\Delta_{kx}}\widetilde{c^{\ast}_{e}}\widetilde{c}_{g} , \\
\frac{d(c_{g}^{\ast}c_{e})}{dt} &= -\frac{\Omega_{1}^{\ast}\Omega_{2}}{\frac{\Gamma}{2}+i\Delta}e^{-i\Delta_{kx}}\widetilde{c^{\ast}_{e}}\widetilde{c}_{e} - \frac{\Omega_{1}^{\ast}\Omega_{2}}{\frac{\Gamma}{2}-i\Delta}e^{i\Delta^{'}_{kx}}\widetilde{c^{\ast}_{g}}\widetilde{c}_{g} - \left[\frac{|\Omega_{1}|^{2}}{\frac{\Gamma}{2}+i\Delta}+\frac{|\Omega_{2}|^{2}}{\frac{\Gamma}{2}-i\Delta} \right]\widetilde{c^{\ast}_{g}}\widetilde{c}_{e}, \\
\frac{d(c_{e}^{\ast}c_{e})}{dt} &= -|\Omega_{2}|^{2}\widetilde{c^{\ast}_{e}}\widetilde{c}_{e}\left[\frac{1}{i\Delta+\frac{\Gamma}{2}}+\frac{1}{\frac{\Gamma}{2}-i\Delta}\right]-\frac{\Omega_{2}^{\ast}\Omega_{1}}{\frac{\Gamma}{2}+i\Delta}e^{-i\Delta^{'}_{kx}}\widetilde{c^{\ast}_{g}}\widetilde{c}_{e}-\frac{\Omega_{1}^{\ast}\Omega_{2}}{\frac{\Gamma}{2}-i\Delta}e^{i\Delta^{'}_{kx}}\widetilde{c^{\ast}_{e}}\widetilde{c}_{g}.
\end{align}
\end{subequations}
\end{widetext}
If we now remember that $\rho_{kj}=c_{k}^{\ast}c_{j}$ we can get everything back in terms of the density matrix. We end up with the same set of differential equations as Eqns. (\ref{rho11dynamics}) - (\ref{rho22dynamics}).

\section{Variance Calculations} \label{appendixB}
\subsection{Acceleration variance density matrix approach}
As seen in the main text, the variance of the acceleration is key to determining the error in the atom interferometer. With the dynamics described in section \ref{closeddensmat}, we can determine the variance strictly in terms of density matrix components. First, let us remind ourselves of the fundamental equation for an atom interferometer to measure acceleration,
\begin{equation}
\frac{|c_{e}(t_{f})|^{2}}{|c_{g}(t_{f})|^2} = \frac{\rho_{ee}}{\rho_{gg}} = b + d \mathrm{tan}^{2}\Phi,
\end{equation}
where $\Phi=\vec{a}T^{2}|k_{\mathrm{eff}}|\mathrm{cos}\theta$. We will again make a simplification to have $\alpha=T^{2}|k_{\mathrm{eff}}|\mathrm{cos}\theta$. We then will take the Taylor series expansion of $\mathrm{tan}^{2}\Phi$ around the expected acceleration $\langle a \rangle$:
\begin{widetext}
\begin{equation}
\mathrm{tan}^2(\alpha a) \simeq \mathrm{tan}^2(\alpha \langle a \rangle) + 2 \alpha [\mathrm{tan}(\alpha \langle a \rangle) + \mathrm{tan}^3(\alpha \langle a \rangle)](a-\langle a \rangle).
\end{equation}
\end{widetext}
Here we have used that up to first-order $a-\langle a \rangle$ will be small. This allows us to write,
\begin{equation}
\frac{\rho_{ee}}{\rho_{gg}} = b + d(a-\langle a \rangle),
\end{equation}
where $b=\mathrm{tan}^2(\alpha \langle a \rangle)$ and $d=2 \alpha [\mathrm{tan}(\alpha \langle a \rangle) + \mathrm{tan}^3(\alpha \langle a \rangle)]$. Therefore, the variances are given by,
\begin{widetext}
\begin{subequations}
\begin{align}
var\left (\frac{\rho_{ee}}{\rho_{gg}} \right) &= var(b + d(a-\langle a \rangle)), \\
&= \langle [b + d(a-\langle a \rangle)]^{2} \rangle - \langle [b + d(a-\langle a \rangle)] \rangle^{2}, \\
&= d^{2}\langle (a-\langle a \rangle)^{2} \rangle + 2d\langle a-\langle a \rangle \rangle, \\
&= d^{2}var(a).
\end{align}
\end{subequations}
\end{widetext}
This comes from the fact that $\langle a-\langle a \rangle \rangle = \langle a \rangle - \langle a \rangle = 0$ and by definition $\langle (a-\langle a \rangle)^{2} \rangle$ is the variance. From here we get,
\begin{equation}
var(a) = \frac{1}{d^{2}}var\left( \frac{\rho_{ee}}{\rho_{gg}} \right).
\end{equation}
Similar to before, we see that we need to know the variance of a ratio. In fact, we are using the variance of the ratio from before just in terms of components of the density matrix. Using Eqn. (\ref{ratioexpect}) we get,
\begin{widetext}
\begin{equation}
var\left (\frac{\rho_{ee}}{\rho_{gg}} \right) \simeq \frac{1}{n}\left[ \frac{var(\rho_{ee})}{\langle \rho_{gg} \rangle^{2}} + \frac{\langle \rho_{ee} \rangle^{2}var(\rho_{gg})}{\langle \rho_{gg} \rangle^{4}} - \frac{2\langle \rho_{ee} \rangle \mathrm{cov}(\rho_{ee},\rho_{gg})}{\langle \rho_{gg} \rangle^{3}} \right].
\end{equation}
\end{widetext}
If we assume level populations are Poisson distributions we will have $var(\rho)= \langle \rho \rangle$ giving us,
\begin{equation}
var\left (\frac{\rho_{ee}}{\rho_{gg}} \right) \simeq \frac{1}{n}\frac{\langle \rho_{ee} \rangle}{\langle \rho_{gg} \rangle^{3}}\left[ \langle \rho_{gg} \rangle + \langle \rho_{ee} \rangle - 2\mathrm{cov}(\rho_{ee},\rho_{gg})\right].
\end{equation}

We will next make the assumption that the error of $\rho$ is equal to its standard deviation (i.e. square root of the variance):
\begin{subequations}
\begin{align}
\rho_{ee} &= \langle \rho_{ee} \rangle \pm \sqrt{\langle \rho_{ee} \rangle}, \\
\rho_{gg} &= \langle \rho_{gg} \rangle \pm \sqrt{\langle \rho_{gg} \rangle}.
\end{align}
\end{subequations}
We can then write the covariance as,
\begin{widetext}
\begin{subequations}
\begin{align}
\mathrm{cov}(\rho_{ee},\rho_{gg}) &= \langle \rho_{ee}\rho_{gg} \rangle - \langle \rho_{ee} \rangle \langle \rho_{gg} \rangle, \\
&= \langle \langle \rho_{ee} \rangle \langle \rho_{gg} \rangle \pm \langle \rho_{gg} \rangle\sqrt{\langle \rho_{ee} \rangle} \pm \langle \rho_{ee} \rangle\sqrt{\langle \rho_{gg} \rangle} \pm \sqrt{\langle \rho_{ee} \rangle\langle \rho_{gg} \rangle}\rangle -\langle \rho_{ee} \rangle\langle \rho_{gg} \rangle.
\end{align}
\end{subequations}
\end{widetext}
The average of the $\pm$ terms will be zero. It can be seen then that the covariance will come out to be zero as well leaving the variance of the ratio to be,
\begin{equation}
var\left (\frac{\rho_{ee}}{\rho_{gg}} \right) \simeq \frac{1}{n}\frac{\langle \rho_{ee} \rangle}{\langle \rho_{gg} \rangle^{3}}\left[ \langle \rho_{gg} \rangle + \langle \rho_{ee} \rangle \right].
\end{equation}
Thus, we get our variance for the acceleration to be
\begin{equation}
\mathrm{var}(a) \simeq \frac{1}{n}\frac{\langle \rho_{ee} \rangle}{\langle \rho_{gg} \rangle^{3}}\frac{\left[ \langle \rho_{gg} \rangle + \langle \rho_{ee} \rangle \right]}{\left (2 \alpha [\mathrm{tan}(\alpha \langle a \rangle) + \mathrm{tan}^3(\alpha \langle a \rangle)] \right)^{2}}
\end{equation}
It is worth noting again, that $\alpha$ will change depending on if you consider a traditional $\pi/2-\pi-\pi/2$ pulse sequence or the multi-Raman pulse sequence.

\subsection{Variance, covariance, and analytical Q}
\subsubsection{Analytical Q}

Now that we have determined the dynamics of our open quantum system, including spontaneous emission, we can study the
effects of quantum information loss due to the higher probability of spontaneous decay from the intermediate high energy state. Let us consider the population of the intermediate excited state a number of excited state lifetimes after a Raman pulse. Our model atom is $^{85}Rb$,which consists of two ground states ($\ket{F=2}$ and $\ket{F=3}$) with corresponding $2F+1$ magnetic sublevels. We assume in our model that the Raman pulse drives the clock transition $\ket{F=2, m_F=0}\rightarrow \ket{F=3, m_F=0}$. The loss will be dominated by decay into the loss state made up of all sublevels that are not part of the clock transition. For a {\em closed} three-level Lambda system, we have the usual population conservation equation
\begin{subequations}
\begin{equation}
    \left\vert c_{g}\right\vert
^{2} + \left\vert c_{e}\right\vert ^{2} + \left\vert c_{i}\right\vert
^{2}=1.
\end{equation}
However, for our open system, we account for the loss of
quantum information by writing
\begin{equation}
   \left\vert c_{g}\right\vert
^{2}+\left\vert c_{e}\right\vert ^{2} + \left\vert c_{i}\right\vert
^{2}=1-Q,
\end{equation}
\end{subequations}
where $Q$ is the amount of quantum
information lost for a given pulse. We can connect the population in the intermediate high energy state to the quantum information loss as follows%
\begin{equation}
Q= {\displaystyle\int\nolimits_{t_{0}}^{t}}\gamma_{l}\widetilde{c}_{i}^{\ast}\widetilde{c}_{i}dt. \label{Qintegral}
\end{equation}
In general, Raman pulses are long relative to the inverse decay rate of the excited state ($\gamma_l$), so we can consider a time long enough so that $\gamma_lt\gg 1$. In this limit, only the oscillatory time-dependent term of
$\widetilde{c}_{i}$ (i.e., the second bracketed term of the right-hand side in equation \eqref{cipopFinalv1}) remains. This gives us a quasi-steady-state quantity $\overline{\widetilde{c_i}}$  as a function of slowly varying $\widetilde{c_g}$ and $\widetilde{c_e}$ 
\begin{widetext}
\begin{align}
\overline{\widetilde{c}_{i}}&=\left[  \frac{-i\Omega_{1}^{\ast}}{-i\Delta
+\frac{\gamma_{l}}{2}}e^{i(\mathbf{k}_{1}\cdot\mathbf{x}_{g}+\Delta t)}%
\widetilde{c}_{g}+\frac{-i\Omega_{2}^{\ast}}{-i(\Delta+\delta)+\frac{\gamma_{l}}%
{2}}e^{i(\mathbf{k}_{2}\cdot\mathbf{x}_{e}+(\Delta+\delta)t)}\widetilde{c}%
_{e}\right]  
\label{Eq:ciSS}
\end{align}
\end{widetext}
Using equation (\ref{Eq:ciSS}) and choosing $\Omega_{1}=\Omega_{2}$ to avoid AC Stark shifts proportional to $|\Omega_1|^2-|\Omega_2|^2$ \cite{PB97}, we obtain
\begin{widetext}
\begin{align}
\overline{\widetilde{c}_{i}^{\ast}} \,\overline{\widetilde{c}_{i}}&\simeq\ \left\vert \Omega_{1}\right\vert ^{2} \left[ \frac{\widetilde{c}_{g}^{2}}{\left(\Delta^{2}+\left(\frac{\gamma_{l}}{2}\right)^{2}\right)}
+\frac{\widetilde{c}_{e}^{2}}{\left((\Delta+\delta)^{2}+\left(\frac{\gamma_{l}}{2}\right)^{2}\right)}+  \left(\frac{\widetilde{c}_{g}\widetilde{c}^{\ast}_{e}}{(-i\Delta+\frac{\gamma_{l}}{2})(i(\Delta+\delta)+\frac{\gamma_{l}}{2})}e^{i(\Delta_{kx}+\delta t)} + c.c. \right) \right]  \label{cici}.
\end{align}
\end{widetext}
Next, we will Taylor expand the terms that will depend on $e^{i\delta t}$ to first order. Plugging Eqn. (\ref{cici}) with the Taylor expanded terms into the integration of Eqn. (\ref{Qintegral}), we will be left with a term linear in $t_{tot}$ (where $t_{tot}=t-t_{0}$) and a term to the second order in $t_{tot}$. If we also consider that $t_{tot}$ will be on the order of microseconds this Taylor expansion will hold. Also, if we consider that for an efficient atom interferometer $\Delta \gg \gamma_{l}$, we can ignore the second order and higher terms. It is also worth noting that we are considering that the detuning $\Delta$ is well within $1/\sigma_{T}$, where $\sigma_{T}$ is the pulse width. Then, we get the following for our quantum information loss%
\begin{widetext}
\begin{equation}
Q=\gamma_{l}\left\vert \Omega_{1}\right\vert ^{2} t_{\mathrm{tot}} \left[ \frac{\widetilde{c}_{g}^{2}}{\left(\Delta^{2}+\left(\frac{\gamma_{l}}{2}\right)^{2}\right)}+\frac{\widetilde{c}_{e}^{2}}{\left((\Delta+\delta)^{2}+\left(\frac{\gamma_{l}}{2}\right)^{2}\right)}+  \left(\frac{\widetilde{c}_{g}\widetilde{c}^{\ast}_{e}}{(-i\Delta+\frac{\gamma_{l}}{2})(i(\Delta+\delta)+\frac{\gamma_{l}}{2})} e^{i\Delta_{kx}} + c.c. \right) \right] \label{finalQ}.
\end{equation}
\end{widetext}

\subsubsection{Variance and covariance relation}
To determine the variance of the quantum information loss, $Q$, we will consider the operator $A$ made up of the jump operators $c^{\dagger}=|i \rangle \langle l|$ and $c=|l \rangle \langle i|$ such that $A=cc^{\dagger}$. The variance of our operator $A$ will be defined by
\begin{align}
\mathrm{var}(A) = \langle A^{2} \rangle - \langle A \rangle^{2}.
\end{align}
We plug in the definition of the expectation value to get
\begin{subequations}
\begin{align}
\mathrm{var}(A) &= \langle \psi|cc^{\dagger}cc^{\dagger} |\psi \rangle - \langle \psi|cc^{\dagger} |\psi \rangle^{2}, \\
\mathrm{var}(A) &= \langle\psi|cc^{\dagger}|\psi\rangle - \langle\psi|cc^{\dagger} |\psi\rangle^{2}.
\end{align}
\end{subequations}
Next we can use that $\langle\psi|cc^{\dagger}|\psi\rangle = \langle\psi|l \rangle \langle l|\psi\rangle = |c_{l}|^{2}$. We have defined the population of the loss state as $Q$ which is defined by Eqn. (\ref{finalQ}). In summary, we have
\begin{equation}
    \mathrm{var}(cc^\dagger) = |c_l|^2 - |c_l|^4 = |c_l|^2(1 - |c_l|^2).
\end{equation}
Note that we proved that $|c_l|^2$ is relatively a constant over the time of interest (see Eqn. (\ref{finalQ})). This leads us to the variance of the quantum information loss per pulse to be 
\begin{subequations}
\begin{align}
\mathrm{var}(Q) &= \gamma_l^2 t^2_\mathrm{tot} \mathrm{var}(cc^\dagger) = \gamma_l^2 t^2_\mathrm{tot} \frac{Q}{\gamma_l t_\mathrm{tot}}\left(1- \frac{Q}{\gamma_l t_\mathrm{tot}} \right) \\
\mathrm{var}(Q) &= \gamma_l t_\mathrm{tot} Q\left( 1 - \frac{Q}{\gamma_l t_\mathrm{tot}}\right)
\end{align}
\end{subequations}
A quick sanity check will show that this will be valid as $Q$ will be less than 1 and $\gamma_{l}t_{tot} \gg 1$.

With the variance of the quantum information loss determined, we can now begin to understand the covariance between the excited state population and the quantum information loss. Let us first consider how we will define the excited state population and quantum information loss. The error in the excited state population will come from two sources, the error in the Raman process measurement and from the spontaneous emission. The error of quantum information loss will be the error in the population of the loss state and will also be due to spontaneous emission. We will treat these errors as being random error sources. We rewrite the excited state population, ground state population, and the quantum information loss as 
\begin{subequations}
\begin{align}
|c_{e}|^{2} &= \overline{|c_{e}|^{2}} + \epsilon_{M} + \epsilon_{\mathrm{\Gamma}} , \\
|c_{g}|^{2} &= \overline{|c_{g}|^{2}} + \epsilon'_{\mathrm{\Gamma}} , \\
Q &= \overline{Q} +  \epsilon_{Q}.
\end{align}
\end{subequations}
Here $\epsilon_{M}$ and $\epsilon_{\mathrm{\Gamma}}$ are the errors of the excited state population due to the measurement process and spontaneous emission respectively, $\epsilon'_{\mathrm{\Gamma}}$ is the error of the ground state population due to spontaneous emission, and $\epsilon_{Q}$ is the error in the quantum information loss due to spontaneous emission.

We can use these rewritten definitions for the excited state population and the quantum information loss to get the covariance as follows
\begin{widetext}
\begin{subequations}
\begin{align}
\mathrm{cov}(|c_{e}|^{2},Q) &= \langle (\overline{|c_{e}|^{2}} + \epsilon_{M} + \epsilon_{\mathrm{\Gamma}})(\overline{Q} +  \epsilon_{Q}) \rangle - \langle \overline{|c_{e}|^{2}} + \epsilon_{M} + \epsilon_{\mathrm{\Gamma}}\rangle \langle \overline{Q} +  \epsilon_{Q}\rangle , \\
\mathrm{cov}(|c_{e}|^{2},Q) &= \langle \epsilon_{\mathrm{\Gamma}}\epsilon_{Q} \rangle .
\end{align}
\end{subequations}
\end{widetext}
We have used here that the only correlated error will be due to the errors from spontaneous emission. Let us now consider the fact that $|c_{e}|^{2}+|c_{g}|^{2}+|c_{l}|^2 = |c_{e}|^{2}+|c_{g}|^{2}+mQ/\gamma_{l} t_\mathrm{tot} = 1$, where $m=2 N_{R}$. The error must have a mean value of zero, and approximately $\epsilon_{\Gamma} \simeq \epsilon'_{\Gamma}$ for many Raman pulses, so we will have $2 \epsilon_\Gamma+m\epsilon_{Q}/\gamma_l t_\mathrm{tot}=0$ while ignoring the measurement error $\epsilon_M$. Then, we will get the final form of the covariance to be
\begin{align}
\mathrm{cov}(|c_{e}|^{2},Q) &= -\frac{m}{2 \gamma_{l} t_\mathrm{tot}}\langle\epsilon_{Q}^{2}\rangle = -\frac{m}{2\gamma_l t_\mathrm{tot}}\mathrm{var}(Q) \nonumber \\
&= - \frac{m}{2} Q \left(1 - \frac{Q}{\gamma_l t_\mathrm{tot}} \right).
\end{align}
\end{appendix}

\end{document}